\documentclass{article}
\usepackage[utf8]{inputenc}
\usepackage[margin=1in]{geometry}
\usepackage{amsfonts, amsmath, amssymb, amsthm, bbm, enumitem, dsfont}
\usepackage[usenames,dvipsnames,table]{xcolor}
\usepackage{preamble}

\newcommand{\fr}{\frac}
\newcommand{\lt}{\left}
\newcommand{\rt}{\right}
\newcommand{\rhat}{\widehat{\rho}}
\newcommand*{\defeq}{\stackrel{\text{def}}{=}}
\newcommand{\Gbar}{\overline{\Gamma}}
\newcommand{\ind}{\mathbbm{1}}

\newcommand{\bbP}{\mathbb{P}}
\newcommand{\bbR}{\mathbb{R}}
\newcommand{\bbC}{\mathbb{C}}
\newcommand{\cA}{\mathcal{A}}
\newcommand{\cE}{\mathcal{E}}
\newcommand{\cM}{\mathcal{M}}
\newcommand{\cN}{\mathcal{N}}
\newcommand{\cS}{\mathcal{S}}
\newcommand{\cT}{\mathcal{T}}
\newcommand{\cZ}{\mathcal{Z}}

\newcommand{\bx}{{\boldsymbol x}}
\newcommand{\blam}{{\boldsymbol \lambda}}
\newcommand{\bmu}{{\boldsymbol \mu}}
\newcommand{\olam}{{\bar \lambda}}
\newcommand{\diag}{\mathrm{diag}}
\newcommand{\unif}{\mathrm{unif}}
\newcommand{\good}{\mathsf{good}}

\newcommand{\op}{\mathsf{op}}
\newcommand{\tr}{\mathsf{tr}}

\newcommand{\Leb}{{\mathsf{Leb}}}

\newcommand{\eps}{\varepsilon}

\newcommand{\GOE}{\mathrm{GOE}}
\newcommand{\GOEc}{\mathrm{GOE}^*}

\newcommand{\de}{{\mathrm{d}}}
\newcommand{\Vol}{\mathrm{Vol}}

\definecolor{Gred}{RGB}{219, 50, 54}
\definecolor{ToCgreen}{RGB}{0, 128, 0}

\usepackage{hyperref}
\hypersetup{
      colorlinks=true,
  citecolor=ToCgreen,
  linkcolor=Sepia,
  filecolor=Gred,
  urlcolor=Gred
  }

\renewcommand{\Pr}{\mathop{\mathbb{P}}}

\newcommand{\sitan}[1]{{\color{red} [sitan: #1]}}

\title{When Does Adaptivity Help for Quantum State Learning?}
\author{
    Sitan Chen \thanks{UC Berkeley. This work was supported in part by NSF Award 2103300.}
    \and
    Brice Huang \thanks{MIT. This work was supported by an NSF Graduate Research Fellowship, a Siebel scholarship, NSF awards CCF-1940205 and DMS-1940092, and NSF-Simons collaboration grant DMS-2031883.}
    \and
    Jerry Li \thanks{Microsoft Research}
    \and
    Allen Liu \thanks{MIT.  This work was supported in part by an NSF Graduate Research Fellowship, a Fannie and John Hertz Foundation Fellowship  and NSF awards CCF-1453261 and CCF-1565235.}
    \and
    Mark Sellke \thanks{Stanford. This work was supported by an NSF Graduate Research Fellowship, a Stanford Graduate Fellowship, and NSF grant CCF-2006489.}
}
\begin{document}
\maketitle

\begin{abstract}
    % We consider the classic question of state tomography: given copies of an unknown quantum state $\rho\in\mathbb{C}^{d\times d}$, output $\widehat{\rho}$ for which $\norm{\rho - \widehat{\rho}}_{\tr} \le \eps$. When one is allowed to make coherent measurements entangled across all copies, $\Theta(d^2/\eps^2)$ copies are necessary and sufficient~\cite{haah2017sample,o2016efficient}. Unfortunately, the protocols achieving this rate incur large quantum memory overheads that preclude implementation on current or near-term devices. On the other hand, the best known protocol using incoherent (single-copy) measurements uses $O(d^3/\eps^2)$ copies~\cite{kueng2017low}, and multiple papers have posed it as an open question to understand whether or not this rate is tight~\cite{haah2017sample,bubeck2020entanglement}. In this work, we fully resolve this question, by showing that any protocol using incoherent measurements, even if they are chosen adaptively, requires $\Omega(d^3/\eps^2)$ copies, matching the upper bound of~\cite{kueng2017low}.
    % We do so by a new proof technique which directly bounds the ``tilt'' of the posterior distribution after measurements, which yields a surprisingly short proof of our lower bound, and which we believe may be of independent interest.
% Likelihood ratios are all you need!
    We consider the classic question of state tomography: given copies of an unknown quantum state $\rho\in\mathbb{C}^{d\times d}$, output $\widehat{\rho}$ which is close to $\rho$ in some sense, e.g. trace distance or fidelity. When one is allowed to make coherent measurements entangled across all copies, $\Theta(d^2/\eps^2)$ copies are necessary and sufficient to get trace distance $\eps$~\cite{haah2017sample,o2016efficient}. Unfortunately, the protocols achieving this rate incur large quantum memory overheads that preclude implementation on near-term devices. On the other hand, the best known protocol using incoherent (single-copy) measurements uses $O(d^3/\eps^2)$ copies~\cite{kueng2017low}, and multiple papers have posed it as an open question to understand whether or not this rate is tight~\cite{haah2017sample,bubeck2020entanglement}. In this work, we fully resolve this question, by showing that any protocol using incoherent measurements, even if they are chosen adaptively, requires $\Omega(d^3/\eps^2)$ copies, matching the upper bound of~\cite{kueng2017low}.
    We do so by a new proof technique which directly bounds the ``tilt'' of the posterior distribution after measurements, which yields a surprisingly short proof of our lower bound, and which we believe may be of independent interest.

While this implies that adaptivity does not help for tomography with respect to trace distance, we show that it actually does help for tomography with respect to infidelity. We give an adaptive algorithm that outputs a state which is $\gamma$-close in infidelity to $\rho$ using only $\wt{O}(d^3/\gamma)$ copies, which is optimal for incoherent measurements. In contrast, it is known~\cite{haah2017sample} that any nonadaptive algorithm requires $\Omega(d^3/\gamma^2)$ copies. While it is folklore that in $2$ dimensions, one can achieve a scaling of $O(1/\gamma)$, to the best of our knowledge, our algorithm is the first to achieve the optimal rate in all dimensions.
\end{abstract}

\section{Introduction}
\label{sec:intro}

In this paper, we consider the problem of \emph{quantum state tomography}.
Here, the algorithm is given $n$ copies of an unknown $d$-dimensional quantum mixed state $\rho$, and the goal is to output $\widehat{\rho}$ which is close to $\rho$ in some sense.\footnote{Here and throughout the introduction, we will consider the setting where the success probability is some fixed constant.}
There are several standard ways to measure closeness used in the literature.
Arguably the two most important and well-studied notions are that of \emph{trace distance} and \emph{infidelity}\footnote{The fidelity between $\rho,\sigma$ is $F(\rho,\sigma) = \tr(\sqrt{\sqrt{\rho}\sigma\sqrt{\rho}})^2$, and the infidelity is $1 - F(\rho,\sigma)$.}, which are the natural quantum analogues of total variation distance and Hellinger distance, respectively.
From a mathematical point of view, tomography is arguably the most fundamental quantum state learning task, and can also be seen as the natural non-commutative generalization of the classical problem of estimating a discrete distribution from samples.
From a practical point of view, state tomography has many applications to the verification of quantum technologies~\cite{banaszek2013focus}.

In the \emph{coherent} setting, that is, when the learner is allowed to make arbitrary measurements to the product state $\rho^{\otimes n}$, the situation is very well-understood. \cite{haah2017sample} demonstrated that for this problem, $n = O (d^2 \log (d / \eps) / \eps^2)$ copies suffice to to learn a state to error $\eps$ in trace distance, and $n = \Omega (d^2 / \eps^2)$ copies are necessary.
Concurrently, \cite{o2016efficient} removed the logarithmic factor in the upper bound, thus establishing that the optimal rate for this problem is $n = \Theta (d^2 / \eps^2)$.
The aforementioned paper of~\cite{haah2017sample} also showed that, up to a single logarithmic factor, $\Omega (d^2 / \gamma)$ copies are sufficient and necessary to learn a state to infidelity $\gamma$.
Tight bounds are also known in many other settings, including for low rank states~\cite{haah2017sample,o2016efficient,o2017efficient}, as well as for a variety of related testing problems~\cite{o2015quantum,buadescu2019quantum}.

However, in virtually all cases, the upper bounds that achieve the optimal rates require heavily entangled measurements.
These measurements require all $n$ copies of $\rho$ to be prepared simultaneously, which---in addition to the complexity of preparing the measurements themselves---renders such approaches currently impractical.
In light of this, there has been a recent surge of interest in the \emph{incoherent} setting.
Here, the algorithm is restricted to only making measurements of a single copy of $\rho$ at a time.
While such measurements are strictly weaker than general entangled measurements, they are much more practical and can be performed on real world quantum computers~\cite{huang2021quantum}.

One of the main difficulties that arises when dealing with incoherent measurements is understanding the power of \emph{adaptivity}.
In general, an algorithm that uses incoherent measurements can sequentially measure each copy of $\rho$, and moreover can choose how to measure the $i$-th copy of $\rho$ based on the results of the previous $i - 1$ measurements.
We say such an algorithm is \emph{adaptive}.
In contrast, a \emph{nonadaptive} algorithm must specify all of its measurements ahead of time.

Clearly, adaptive measurements are strictly more general than nonadaptive ones. On the other hand, to date, it has been unclear whether or not adaptivity actually yields better rates for any natural quantum learning problems.
In large part, this is because until recently, it was not known how to obtain tight lower bounds against adaptive algorithms.
Existing lower bound frameworks in the literature could only prove lower bounds against nonadaptive measurements; such algorithms are significantly easier to work with, and existing techniques from the classical learning literature could be adapted to the quantum setting.

However, starting with recent work of~\cite{bubeck2020entanglement}, there have been a flurry of works giving lower bounds against adaptive algorithms for a number of natural quantum learning problems.
Now, lower bounds are known for a number of testing problems, such as
mixedness testing and identity testing~\cite{bubeck2020entanglement,chen2021toward,chen2022tight}, shadow tomography~\cite{huang2021information,chen2022exponential,chen2021hierarchy}, purity testing~\cite{aharonov2022quantum,chen2022exponential}.
In all of these cases, the lower bound against adaptive algorithms matches the best nonadaptive bound (up to polylogarithmic factors), so for all of these problems, adaptivity is not necessary to achieve optimal rates.

In contrast, for state tomography, progress has been slow in understanding the power of adaptivity.
It is known that nonadaptively, $\Theta (d^3 / \eps^2)$ copies are necessary and sufficient for learning to error $\eps$ in trace distance~\cite{kueng2017low,haah2017sample}, and $\Theta (d^3 / \gamma^2)$ copies are necessary and sufficient to learn to infidelity $\gamma$~\cite{haah2017sample}.
Similarly, as mentioned above, tight rates are known for the fully coherent setting (up to log factors).
But nothing nontrivial is known in the adaptive setting: prior to our work, we did not know any adaptive algorithms that improve upon the best known nonadaptive rates, nor did we know any lower bounds against adaptive algorithms which were better than those known against coherent algorithms.
Indeed, understanding the power of adaptive but incoherent measurements for quantum tomography has been posed as an open question by multiple papers~\cite{haah2017sample,bubeck2020entanglement}.

This problem is interesting on both fronts.
From the lower bounds perspective, it seems that fundamentally new ideas are necessary to develop lower bounds for tomography.
This is because all previous lower bound techniques for adaptive algorithms only held for testing problems, or via reductions to testing problems.
These techniques will consequently fail for tomography, which is fundamentally a learning problem.
From the upper bounds perspective, tomography in infidelity is one of the few natural candidates for a quantum learning problem where adaptive algorithms are provably better than nonadaptive ones.
Indeed, it is folklore that in 2 dimensions, there are adaptive algorithms which achieve a scaling of $O(1 / \gamma)$, although we could not find a reference to a full proof of this fact.

\paragraph{Our Results.}
In this paper, we fully resolve the copy complexity of quantum state tomography with (potentially adaptively chosen) incoherent measurements, for both learning in trace distance and learning in infidelity.
For trace distance, our main result is the following:
\begin{theorem}[Informal, see Theorem~\ref{thm:tomo_lbd}]\label{thm:tomo_informal}
     With incoherent measurements, $\Omega(d^3/\eps^2)$ copies are necessary for tomography of $d$-dimensional states to $\eps$ error in trace distance.
\end{theorem}
\noindent
This matches the upper bound given by nonadaptive algorithms, and therefore for this problem, adaptivity provably does not help.
In contrast, for learning in infidelity, our main result is the following:

\begin{theorem}[Informal, see Theorem~\ref{thm:learn-in-fidelity}]\label{thm:tomo_upper_informal}
     With adaptive, incoherent measurements, $\widetilde{O}(d^3/\gamma)$ copies\footnote{We use $f = \wt{O}(g)$ to denote that $f = C\cdot g\cdot \log^c(g))$ for some absolute constants $c, C > 0$.} are sufficient for tomography of $d$-dimensional states to $\gamma$ infidelity. 
\end{theorem}
\noindent
Because trace distance and fidelity are related via $1 - F(\rho,\sigma) \le \norm{\rho - \sigma}_{\tr} \le \sqrt{2(1 - F(\rho,\sigma))}$ \cite{fuchs1999cryptographic}, Theorem~\ref{thm:tomo_informal} implies that $\Omega(d^3/\gamma)$ copies are \emph{necessary} for tomography in infidelity, so up to polylogarithmic factors, our upper bound in Theorem~\ref{thm:tomo_upper_informal} settles the complexity of tomography in infidelity with incoherent measurements.
In contrast, \cite{haah2017sample} showed that $\Omega(d^3/\gamma^2)$ copies are necessary with nonadaptive measurements, so the rate in Theorem~\ref{thm:tomo_upper_informal} is provably better than what can be achieved with nonadaptive measurements, and thus provides arguably the first natural instance of a separation between the power of adaptive versus nonadaptive measurements.

\paragraph{Our Techniques.}
Here, we give a very high level discussion of our proof techniques for Theorems~\ref{thm:tomo_informal} and~\ref{thm:tomo_upper_informal}.
For a more detailed discussion, see Section~\ref{sec:overview}.

We begin with the lower bound.
The main difficulty with proving lower bounds for state tomography with incoherent measurements is that essentially all the existing lower bound frameworks in the literature were fundamentally for testing problems.
In such settings, it suffices to demonstrate hardness for a point-versus-mixture distinguishing task, where the goal is to distinguish between the case where the unknown mixed state is a single point versus the case where it is drawn from a mixture over alternate hypotheses.
Such a setup is mathematically nice because the resulting likelihood ratios have a (relatively) simple multilinear form.
However, for learning tasks, no such reduction exists; indeed, it is more natural to demonstrate hardness for a mixture-versus-mixture distinguishing task, but here the resulting likelihood ratios are much more complicated.
Indeed, this phenomena seems to appear more generally in a variety of other (classical) learning settings, see e.g.~\cite{schramm2020computational}.

We avoid this by directly bounding how much information the algorithm can learn from incoherent measurements, a technique which we believe may be of independent interest.
We  demonstrate that, for a carefully chosen prior on mixed states, the posterior distribution of the algorithm after $o (d^3 / \eps^2)$ incoherent measurements is anti-concentrated around the true mixed state.
It is perhaps surprising that we are able to directly bound the behavior of the posterior distribution, but it turns out that the ``tilt'' caused by the measurements can be upper bounded ``by hand'', and then the required anti-concentration follows from classical results in random matrix theory.
% An additionally nice feature of this approach is that the resulting proof is incredibly short---indeed, the entire proof fits in essentially 4 pages!

We now turn our attention to the upper bound.
First, let us briefly discuss why adaptivity may help for learning in infidelity.
At a very high level, learning in infidelity seems to correspond to learning the eigenvalues and eigenvectors of the density matrix to some degree of relative accuracy.
However, nonadaptive algorithms are unable to do this easily: the small eigenspaces are hidden by the ``noise'' caused by the large eigenvalues.
One the other hand, adaptive algorithms can learn the large eigenspaces, then project them away to reveal the information about the smaller eigenvalues.

The main challenge with this approach is dealing with the error accumulation in the infidelity as we iterate this process.
In particular, we cannot exactly learn the large eigenspaces, and so we must control the error the projections incur.
In constant dimensions, one can directly brute force the calculations, but in high dimensions, the calculations become intractable.
Our main technical contribution for this part of the paper is a new technique to bound the fidelity between two noncommuting states, by carefully guessing a matrix square root for their symmetrized product.

\paragraph{Comparison to Prior Work.} As discussed above, our techniques are very different from the ones used to prove lower bounds for quantum \emph{testing} in the incoherent measurements setting \cite{bubeck2020entanglement,chen2021toward,chen2022tight,huang2021information,chen2022exponential,chen2021hierarchy,aharonov2022quantum,fawzi2023quantum}. 
Here we briefly mention the sole commonalities between our proof and these works.

First, we use the helpful abstraction of adaptive algorithms as given by decision trees. This is common to all aforementioned works on incoherent measurements, but here we clarify a common misconception: the tree representation is not so much a central technical ingredient as it is a notational necessity for keeping track of the internal state of the algorithm.

Secondly, similar to \cite{chen2022tight}, in place of an ensemble based on Haar-random unitaries, we work with a suitable Gaussian approximation. While the former is the construction of choice for many known quantum property testing lower bounds, it was observed in \cite{chen2022tight} that the latter is more amenable to certain high-degree moment calculations that arise in the proofs of these lower bounds. We emphasize however that such moment calculations are not relevant to the current work, and the Gaussian ensemble turns out to be convenient for fairly different reasons, namely certain density and isotropy properties (see Section~\ref{sec:lbd_overview}).

The similarities between our techniques and previous work essentially end here.
In particular, our main lower bound strategy (which allows us to directly reason about the change in the posterior distribution) is to our knowledge completely novel, and we believe it may be of independent interest.

\paragraph{Additional Related Work.}
% (coherent upper bounds for tomography in trace distance, uses representation-theoretic tools from \cite{alicki1988symmetry,keyl2005estimating}, see also \cite{o2017efficient} for upper bounds under other metrics), coherent lower bounds follow from Holevo's and existence of certain packings of states (see e.g. Theorem 3 in \cite{haah2017sample}), (coherent upper and lower bounds for testing), \cite{aaronson2019shadow,aaronson2018online,aaronson2019gentle,buadescu2021improved} (coherent upper bounds for shadow tomography)

% In addition to the aforementioned upper and lower bounds for state tomography \cite{haah2017sample,kueng2017low}, there has been a flurry of recent work on various quantum testing and estimation tasks in the context of incoherent measurements. For mixedness testing, the quantum analogue of classical uniformity testing, the copy complexity is $\Theta(d/\eps^2)$ with coherent measurements \cite{o2015quantum} but $\Theta(d^{3/2}/\eps^2)$ with incoherent measurements \cite{bubeck2020entanglement,chen2022tight} (see also \cite{chen2021toward,chen2022tight} for related bounds for state certification). For shadow tomography and related tasks, \cite{huang2021information,chen2022exponential,chen2021hierarchy,huang2021quantum} have shown exponential separations between algorithms that can make coherent measurements versus those that can make incoherent measurements or, more generally, have bounded quantum memory. \cite{aharonov2022quantum,chen2022exponential,huang2021quantum} have shown similar exponential separations for various channel distinguishing tasks.

Apart from the aforementioned bounds for state tomography and quantum state testing, there have also been lower bounds in the incoherent setting when the measurements are partially adaptive or come from a set of bounded size~\cite{lowe2021learning}, and when the measurements are Pauli~\cite{flammia2012quantum}.

We also note there is a large literature on understanding the power that adaptivity affords for tomography in infidelity in the \emph{asymptotic} setting~\cite{bagan2006optimal,hayashi2008asymptotic,mahler2013adaptive,ferrie2015minimax}. They obtain rates that are linear in $1/\gamma$, but unlike our Theorem~\ref{thm:tomo_upper_informal}, these results get some unspecified dependence on $d$ and/or only apply to the regime of $d = O(1)$.

\cite{van2023quantum} gives upper and lower bounds for \emph{pure state} tomography in a different setting where instead of getting copies of the unknown state, one has access to a unitary which prepares the state.

Finally, the recent work of \cite{fawzi:hal-04107265} gives a constant-factor separation between non-adaptive and adaptive algorithms for quantum hypothesis selection, as well as a polynomial separation for a problem in which one is promised that an unknown state can be diagonalized in one of $m$ known bases and would like to approximate the state in trace distance by measuring copies. Their separation is with respect to the parameters $d$ and $m$, rather than $\epsilon$ as in our work.

% \TODO{other refs: tomography with state preparation oracle}

% \TODO{change $\delta$ for infidelity to $\gamma$?}

% \allen{Have we defined fidelity yet up to this point?}

% \paragraph{Coherent measurements.}

% {\color{red} TODO (sitan): mention in some conclusion section that we don't get the rank-$r$ case, leave as open question}

\paragraph{Concurrent work.} Our proof of Theorem~\ref{thm:tomo_upper_informal} directly generalizes to give that  $\widetilde{O}(dr^2/\gamma)$ measurements are sufficient when the unknown state has rank-$r$.  In concurrent and independent work, Flammia and O’Donnell~\cite{steveryan} also obtain, up to polylogarithmic factors, an upper bound of $O(dr^2 / \gamma)$ for tomography with adaptive single-copy measurements. Their guarantee is somewhat stronger as their error is measured in quantum relative entropy, which is an upper bound on infidelity. They also obtain a bound of $\wt{O}(d^{3/2} r^{3/2} / \gamma)$ for the same problem where $\gamma$ is given by Bures chi-squared divergence, which also upper bounds infidelity. % They apply this result to give an algorithm for testing the mutual information in bipartite quantum states.
% \ms{I just skimmed but was confused by this paragraph. Might it make sense to say: concurrently, and relatedly to Theorem~\ref{thm:tomo_upper_informal}, \cite{steveryan} proved an upper bound of $O(dr^2 / \gamma)$ for tomography with adaptive single-copy measurements where $r$ is the rank.}

\section{Technical Overview}
\label{sec:overview}

\paragraph{Notation.} In addition to standard big-O notation, we sometimes also use the notation $f\lesssim g$ (resp. $f\gtrsim g$) to denote that there exists an absolute constant $C > 0$ such that $f \le Cg$ (resp. $f \ge Cg$).

\subsection{Preliminaries}

Throughout, let $\rho$ denote a quantum state.
We recall that a quantum state is a psd Hermitian matrix in $\bbC^{d\times d}$ with trace $1$.

\paragraph{Measurements.} We now define the standard measurement formalism, which is the way algorithms are allowed to interact with a quantum state $\rho$.
\begin{definition}[Positive operator valued measurement (POVM), see e.g.~\cite{nielsen2002quantum}]
A positive operator valued measurement $\cM$ is a finite collection of psd matrices $\cM = \{M_z\}_{z \in \cZ}$ satisfying $\sum_z M_z = I_d$.
When a state $\rho$ is measured using $\cM$, we get a draw from a classical distribution over $\cZ$, where we observe $z$ with probability $\tr (\rho M_z)$.
Afterwards, the quantum state is destroyed.
\end{definition}

\noindent In this work we assume that all POVMs used are rank-1. It is a standard fact that this is without loss of generality (see e.g. \cite[Lemma 4.8]{chen2022exponential}). We will sometimes represent a sequence of $n$ measurement outcomes by $\bx = (x_1,\ldots,x_n)$ to denote that in the $i$-th step, the outcome that was observed corresponds to a POVM element which is a scalar multiple of $x_ix_i^\dagger$.

\paragraph{Incoherent Measurements.}
% Next, we formally define what we mean by an algorithm that uses incoherent measurements.
An algorithm using incoherent measurements operates as follows: given $n$ copies of $\rho$, it iteratively measures the $i$-th copy using a POVM (which could depend on the results of previous measurements), records the outcome, and then repeats this process on the $(i + 1)$-th copy.
After having performed all $n$ measurements, it must output an estimate of $\rho$ based on the (classical) sequence of outcomes it has received.
Formally, such an algorithm can be represented as a tree, see Definition~\ref{def:tree}.

% \allen{Should we put stuff below here in the next section i.e. separate the technical stuff for the lower and upper bounds?}

% \TODO{emphasize somewhere this is not the key idea...}

\subsection{Lower Bound}
\label{sec:lbd_overview}

We begin with a heuristic calculation that explains how the threshold $d^3/\eps^2$ enters our lower bound.
We assume heuristically that we have a discrete set $\cS$ of quantum states $\rho \in \bbR^{d\times d}$ satisfying the following properties. 
\begin{enumerate}[label=\alph*), ref=\alph*]
    \item \label{itm:mixed} \emph{Near-mixedness}: for all $\rho \in \cS$, all eigenvalues of $\rho$ lie in $[0.9/d, 1.1/d]$. 
    \item \label{itm:pack} \emph{Packing}: for all $\rho,\rho' \in \cS$, $\norm{\rho-\rho'}_{\tr} \ge 2\eps$. 
    \item \label{itm:dense} \emph{Density}: for a $1-o(1)$ fraction of $\rho_0 \in \cS$, the neighborhood 
    \[
        N(\rho_0) = \lt\{\rho \in \cS : \norm{\rho-\rho_0}_{\tr} \le 100\eps\rt\}
    \]
    has cardinality $|N(\rho_0)| \ge \exp(\Omega(d^2))$.
    \item \label{itm:isotropy} \emph{Isotropy}: for a $1-o(1)$ fraction of $\rho_0 \in \cS$, the distribution $\gamma(\rho_0) = \unif(N(\rho_0))$ is isotropic around $\rho_0$, in the sense that 
    \begin{align}
        \label{eq:isotropy-mean}
        \E_{\rho \sim \gamma(\rho_0)} [\rho] &= \rho_0, \\
        \label{eq:isotropy-variance}
        \E_{\rho \sim \gamma(\rho_0)} [(v^\dagger (\rho-\rho_0) v)^2] &\lesssim \eps^2/d^3, \qquad \forall \norm{v}_2 = 1.
    \end{align}
    (The below argument still works if \eqref{eq:isotropy-mean} holds only approximately, but having it hold with equality simplifies the discussion.)
\end{enumerate}
We expect this can be achieved by a random construction.
The scale $\exp(\Omega(d^2))$ arises from the dimension of the ambient space of quantum states, and the right-hand side $\eps^2/d^3$ of \eqref{eq:isotropy-variance} is the scale we get if we replace $\gamma(\rho_0)$ with an isotropic distribution around $\rho_0$ with appropriate trace distance, for example that of $\rho = \rho_0 + \fr{100\eps}{d}U^\dagger Z U$ for Haar random $U \in \bbR^{d\times d}$ and $Z = \diag(1,\ldots,-1,\ldots)$.
We expect the density and isotropy conditions to hold for all $\rho_0 \in \cS$ not too close to the boundary of the point cloud $\cS$.

Assuming such $\cS$ exists, we will show that it is not possible to learn $\rho_0$ sampled from the hard distribution $\mu = \unif(\cS)$.
Because $\cS$ is a $2\eps$-packing in trace distance, learning $\rho_0 \in \cS$ to $\eps$ in trace distance amounts to recovering $\rho_0$.

The central idea of the proof is that, if $\rho_0$ satisfies the mixedness and isotropy conditions \eqref{itm:mixed}, \eqref{itm:isotropy}, each measurement improves the log likelihood ratio of $\rho_0$ to $N(\rho_0)$ by at most $O(\eps^2/d)$.
So, if $\rho_0$ satisfies the density condition \eqref{itm:dense}, after observing $n \ll d^3/\eps^2$ measurements the improvement $\exp(O(n\eps^2/d))$ in the likelihood ratio is not enough to overcome the cardinality $\exp(\Omega(d^2))$ of $N(\rho_0)$.
Thus the posterior distribution of $\rho_0$ still places much more mass on $N(\rho_0)$ than $\rho_0$ itself.

More concretely, let the observations (from adaptive POVMs) of $\rho_0$ be $\bx = (x_1,\ldots,x_n)$ and denote the posterior distribution of $\rho_0$ by $\nu_\bx(\cdot)$.
Then,
\begin{align}
    \notag
    \fr{\nu_\bx(N(\rho_0))}{\nu_\bx(\rho_0)}
    &= \sum_{\rho \in N(\rho_0)} 
    \prod_{i=1}^n \fr{x_i^\dagger \rho x_i}{x_i^\dagger \rho_0 x_i}
    = |N(\rho_0)| \E_{\rho \sim \gamma(\rho_0)} 
    \prod_{i=1}^n \fr{x_i^\dagger \rho x_i}{x_i^\dagger \rho_0 x_i} \\
    \label{eq:toy-calculation-likelihood-ratio}
    &\ge 
    \exp\lt\{
        \Omega(d^2) + \sum_{i=1}^n \E_{\rho \sim \gamma(\rho_0)} \log \lt(
            1 + \fr{x_i^\dagger (\rho-\rho_0) x_i}{x_i^\dagger \rho_0 x_i}
        \rt)
    \rt\},
\end{align}
where the last step uses Jensen's inequality.
% We assume that most $\rho_0 \in \cS$ satisfy the isotropy conditions
For suitably small $\eps$, Taylor expanding $\log(1+x)$ gives
\begin{align*}
    \E_{\rho \sim \gamma(\rho_0)} \log \lt(
        1 + \fr{x_i^\dagger (\rho-\rho_0) x_i}{x_i^\dagger \rho_0 x_i}
    \rt)
    &\ge 
    \E_{\rho \sim \gamma(\rho_0)}
    \fr{x_i^\dagger (\rho-\rho_0) x_i}{x_i^\dagger \rho_0 x_i}
    - \E_{\rho \sim \gamma(\rho_0)}
    \fr23 \lt(\fr{x_i^\dagger (\rho-\rho_0) x_i}{x_i^\dagger \rho_0 x_i}\rt)^2 && \\
    &\ge 
    - d^2 \E_{\rho \sim \gamma(\rho_0)} [(x_i^\dagger (\rho-\rho_0) x_i)^2]
    && \text{(by \eqref{itm:mixed})} \\
    &\gtrsim -\eps^2/d
    && \text{(by \eqref{eq:isotropy-variance})}.
\end{align*}
The hypothesis $n \ll d^3/\eps^2$ implies $n\eps^2/d \ll d^2$, so \eqref{eq:toy-calculation-likelihood-ratio} is $\exp(\Omega(d^2))$.
Thus the posterior mass on $\rho_0$ is
\[
    \nu_\bx(\rho_0) \le \fr{\nu_\bx(\rho_0)}{\nu_\bx(N(\rho_0))} \le \exp(-\Omega(d^2))
\]
and learning $\rho_0$ is impossible.

However, it is hard to make the above approach rigorous, due to the difficulty of constructing a point set $\cS$ with the required density and isotropy properties.
We sidestep this issue by instead working with a continuous prior $\mu$, namely a perturbation of the maximally mixed state $\fr{1}{d}I_d$ by a suitable trace-centered Gaussian matrix.
In this approach, in lieu of showing $\nu_\bx(N(\rho_0))/\nu_\bx(\rho_0)$ is large, our goal will be to show $\nu_\bx(B(\rho_0,C\eps))/\nu_\bx(B(\rho_0,\eps))$ is large, where $B(\rho_0,\eps)$ denotes a trace distance ball of radius $\eps$ and $C$ is a large constant.
In the new proof, $\gamma(\rho_0)$ will be $\mu$ restricted to $B(\rho,C\eps)$, then \textbf{sub-sampled} to be transparently isotropic.
Being able to sub-sample a measure in this way is a key advantage of the continuous setup.

The central idea of the proof is still that each measurement improves the log likelihood ratio by  $O(\eps^2/d)$, which is not enough to overcome the volume ratio $|B(\rho_0,C\eps)|/|B(\rho_0,\eps)| = C^{\Theta(d^2)}$, which serves as the continuous analogue of the cardinality of $N(\rho_0)$ from the earlier argument.
Although sub-sampling incurs factors that decrease our estimate of $\nu_\bx(B(\rho_0,C\eps))/\nu_\bx(B(\rho_0,\eps))$ by a $\exp(\Omega(d^2))$ factor, this is overcome by setting $C$ to be a suitably large constant. %(we show $C=e^{20}$ is sufficient).

\subsection{Upper Bound}

The first observation about infidelity compared to trace distance is that infidelity is much more sensitive to errors on smaller eigenvalues.  For instance, consider two states $\rho = \text{diag}(\rho_1, \dots , \rho_d)$ and $\rho' = \text{diag}(\rho_1', \dots , \rho_d')$ where $\rho_i' = \rho_i + \Delta_i$ for all $i \in [d]$.  When the $\Delta_i$ are sufficiently small (to ensure the following power series expansion is a good approximation), we have
\[
1 - F(\rho, \rho') = 1 - \sum_{i = 1}^d \sqrt{\rho_i ( \rho_i + \Delta_i)} \approx 1 - \sum_{i = 1}^d \left( \rho_i - \frac{\Delta_i}{2} - \frac{\Delta_i^2}{4\rho_i} \right) = \sum_{i = 1}^d \frac{\Delta_i^2}{4\rho_i} \,.
\]
The $\rho_i$ in the denominator means that it is necessary to get better accuracy on smaller eigenvalues.  Heuristically, this means to get infidelity $\gamma$ we should need to learn eigenvalues of size $\sim \sigma$ to accuracy $\sqrt{\gamma \sigma /d}$.

In the classical setting of estimating a discrete distribution, if we simply take the empirical point estimates, we naturally get finer additive accuracy for the lower probability points.  However, in the quantum setting this is not the case because we do not know the right basis to measure in.  If we run a nonadaptive tomography  algorithm on an unknown state $\rho$ with $d^3/\gamma$ samples, we can (see Theorem~\ref{thm:spectral-estimator}) learn a state $\hat{\rho}$ such that $\norm{\rho - \hat{\rho}}_{\op} \leq \frac{\sqrt{\gamma}}{d}$.  This would suffice for learning to infidelity $\gamma$ if the eigenvalues of $\rho$ are all $\Omega(1/d)$.  However, if $\rho$ has small eigenvalues, they can be dominated by the estimation error in $\hat{\rho}$.  

Our learning algorithm overcomes this issue by running logarithmically many rounds.  In each round, we run a standard tomography algorithm and argue that this accurately learns the large eigenvalues and corresponding eigenspace.  Then we can essentially project out this eigenspace and restrict to the orthogonal complement in all future measurements using appropriately designed POVMs (see Definition~\ref{def:projected-POVM}).  This lets us focus on the smaller eigenvalues and learn those to finer accuracy.  We can then repeat by running a tomography algorithm on the orthogonal complement, learning and projecting out the largest remaining eigenvalues and so on.  At the end of the algorithm we will have learned projection matrices $\Pi_1 = B_1B_1^\top  , \dots , \Pi_t = B_t B_t^\top$ (where $B_i$ has columns forming an orthonormal basis of the corresponding subspace) where $t \sim \log(d/\eps)$ and an estimator $\hat{\rho}$ that is diagonal in the basis given by $\{B_1, \dots , B_t \}$.  Let $d_i$ be the dimension of $B_i$ so $d = d_1 + \dots + d_t$.  In this basis, we can write
\[
\hat{\rho} = \begin{pmatrix}
    \widehat{\rho_1} & & &\\
    & \widehat{\rho_2} & &\\
    & & \ddots &\\
    & & & \widehat{\rho_{t}} \\
\end{pmatrix} \;\;\text{and} \;\; 
\rho = \begin{pmatrix}
    & \rho_1 & \cellcolor{green} & \cellcolor{green} E_1 & \cellcolor{green}  & \\
    & \cellcolor{green} & \rho_2 & \cellcolor{cyan} \hspace{10pt} E_2 & \cellcolor{cyan}  & \\
    & \cellcolor{green} E_1^\top & \cellcolor{cyan} E_2^\top & \ddots & & \\
    & \cellcolor{green} & \cellcolor{cyan} & & \rho_{t}  & \\
\end{pmatrix}
\]
where $\rho_i, \widehat{\rho_i} \in \R^{d_i \times d_i}$, $E_i \in \R^{d_i \times (d_{i+1} + \dots + d_t)}$.  The guarantees of our tomography algorithm give us appropriate scale-sensitive bounds on $\norm{\widehat{\rho_i} - \rho_i}_{\op}$ and $\norm{E_i}_{\op}$. We can then bound the infidelity between $\rho, \hat{\rho}$ in two stages.  We first bound the infidelity coming from the differences on the diagonal blocks i.e. $1 - F(\hat{\rho}, \rho_{\diag})$ where $\rho_{\diag} = \diag(\rho_1, \dots , \rho_t)$ (see Lemma~\ref{lem:on-diag}).  We then bound the infidelity coming from the off-diagonal blocks (see Lemma~\ref{lem:off-diag})  This is the main technical component of the proof.  The matrix square root that appears in the expression for fidelity can become very complicated and unwieldy.  Instead, we bound it by carefully guessing a matrix that lower bounds the square root (see Lemma~\ref{lem:sqrt}).

\section{Proof of Lower Bound}

In this section we prove Theorem~\ref{thm:tomo_lbd}, our hardness result for tomography in trace distance.

\subsection{Lower Bound Framework}

We begin by reviewing a standard framework for representing an adaptive algorithm as a tree.

\begin{definition}[Tree representation, see e.g. \cite{chen2022exponential}]\label{def:tree}
    Fix an unknown $d$-dimensional mixed state $\rho$. An algorithm for state tomography that only uses $n$ incoherent, possibly adaptive, measurements of $\rho$ can be expressed as a pair $(\cT,\cA)$, where $\cT$ is a rooted tree $\cT$ of depth $n$ satisfying the following properties:
    \begin{itemize}[leftmargin=*,itemsep=0pt]
        \item Each node is labeled by a string of vectors $\bx = (x_1,\ldots,x_t)$, where each $x_i$ corresponds to measurement outcome observed in the $i$-th step.
        \item Each node $\bx$ is associated with a probability $p^{\rho}(\bx)$ corresponding to the probability of observing $\bx$ over the course of the algorithm. The probability for the root is 1.
        \item At each non-leaf node, we measure $\rho$ using a rank-1 POVM $\lt\{\omega_x d \cdot xx^{\dagger}\rt\}_x$ to obtain classical outcome $x\in\mathbb{S}^{d-1}$. The children of $\bx$ consist of all strings $\bx' = (x_1,\ldots,x_t,x)$ for which $x$ is a possible POVM outcome.
        \item If $\bx' = (x_1,\ldots,x_t,x)$ is a child of $\bx$, then
        \begin{equation}
            p^{\rho}(\bx') = p^{\rho}(\bx)\cdot \omega_x d \cdot x^{\dagger} \rho x\,.
        \end{equation}
        \item Every root-to-leaf path is length-$n$. Note that $\cT$ and $\rho$ induce a distribution over the leaves of $\cT$.
    \end{itemize}
    $\cA$ is a randomized algorithm that takes as input any leaf $\bx$ of $\cT$ and outputs a state $\cA(\bx)$. The output of $(\cT,\cA)$ upon measuring $n$ copies of a state $\rho$ is the random variable $\cA(\bx)$, where $\bx$ is sampled from the aforementioned distribution over the leaves of $\cT$.
\end{definition}
\noindent

We also recall the definition of the Gaussian Orthogonal Ensemble (GOE) and define a trace-centered variant, which will be the basis of our hard distribution.

\begin{definition}[GOE, Trace-centered GOE]
    A sample $G \sim \GOE(d)$ is a symmetric matrix with independent Gaussians on and above the diagonal, with $G_{i,i} \sim \cN(0,2/d)$ and $G_{i,j} \sim \cN(0,1/d)$ for $i<j$.
    A sample $G' \sim \GOEc(d)$ is sampled by $G' = G - \Tr(G)$ where $G \sim \GOE(d)$.
\end{definition}

The trace-centered GOE also features in the hard instance of \cite{chen2022tight} for state certification, but as discussed in Section~\ref{sec:intro}, our reason for working with this ensemble is very different from that of~\cite{chen2022tight}.
We recall the following standard fact about extremal eigenvalues of the GOE matrix.

\begin{lemma}[{\cite[Lemma 6.2]{chen2022exponential}}]
    \label{lem:goe-op-norm}
    If $G \sim \GOEc(d)$, then $\norm{G}_\op \le 3$ with probability $1-e^{-\Omega(d)}$.
\end{lemma}

\subsection{Construction of Hard Distribution}

We construct the following hard distribution $\mu$ of quantum states.
Let $U \subseteq \bbR^{d\times d}$ be the affine subspace of symmetric matrices with trace $1$ and $U_0 \subseteq \bbR^{d\times d}$ be the linear subspace of symmetric matrices with trace $0$.
These spaces inherit the inner product of $\bbR^{d\times d}$, which defines Lebesgue measures $\Leb_U$ and $\Leb_{U_0}$ on them.
Let $\sigma = \fr{1}{100}$ be a small constant.
A sample $\rho \sim \mu$ is generated by
\[
    \rho = \fr1d (I_d + \sigma G),
\]
where $G$ is a sample from $\GOEc(d)$ conditioned on $\norm{G}_\op \le 4$.
Note that such matrices are clearly psd and thus valid quantum states.
Concretely, $\mu$ has density (with respect to $\Leb_U$)
\[
    \mu(\rho)
    =
    \fr{1}{Z}
    \exp\lt(
        -\fr{d^3}{4\sigma^2}
        \norm{\rho - \fr{1}{d}I_d}_F^2
    \rt)
    \ind\{\rho \in S_{\supp}\},
    % \ind\lt\{
    %     \norm{\rho - \fr{1}{d}I_d}_\op
    %     \le
    %     \fr{4\sigma}{d}
    % \rt\},
    \qquad
    S_{\supp} = \lt\{
        \rho \in U : \norm{\rho - \fr{1}{d}I_d}_{\op} \le \fr{4\sigma}{d}
    \rt\}.
\]
where $Z$ is a normalizing constant.
Further define a set of ``good" states
\[
    % S_{\all} = \lt\{
    %     \rho \in U : \norm{\rho - \fr{1}{d}I_d}_{\op} \le \fr{4\sigma}{d}
    % \rt\},
    % \qquad
    S_{\good} = \lt\{
        \rho \in U : \norm{\rho - \fr{1}{d}I_d}_{\op} \le \fr{3\sigma}{d}
    \rt\},
\]
which corresponds to the event
% These correspond to the events $\norm{G}_\op \le 4$ and
$\norm{G}_\op \le 3$.
Due to Lemma~\ref{lem:goe-op-norm}, $\mu(S_{\good}) \ge 1-e^{-\Omega(d)}$.

In the below proof, we will show that all $\rho_0 \in S_{\good}$ are hard to learn.
The important property of $S_{\good}$ is that it is far from the boundary of $\supp(\mu) = S_{\supp}$; this  ensures that we can choose a suitable sub-sampling of $\mu$ in a neighborhood of $\rho_0$, which is isotropic around $\rho_0$.

Finally we record the following straightforward fact.
\begin{lemma}
    \label{lem:mu-ratio}
    For all $\rho,\rho' \in S_{\supp}$, $\exp(-4d^2) \le \mu(\rho)/\mu(\rho') \le \exp(4d^2)$.
\end{lemma}
\begin{proof}
    For all $\rho \in S_{\supp}$,
    \[
        0\le
        \fr{d^3}{4\sigma^2} \norm{\rho-\fr{1}{d}I}_F^2
        \le
        \fr{d^4}{4\sigma^2}\norm{\rho-\fr{1}{d}I}_{\op}^2
        \le
        4d^2. \qedhere
    \]
\end{proof}

\subsection{Anticoncentration of Posterior Distribution}

Fix a tomography algorithm $(\cT,\cA)$ as in Definition~\ref{def:tree}, and let $\cT_\rho$ denote the distribution over observation sequences $\bx = (x_1,\ldots,x_n)$ when $\cT$ is run on state $\rho$.
Note that for any states $\rho, \rho'$ in the support of $\mu$, the likelihood ratio % Radon-Nikodym derivative (likelihood ratio)
\[
    \fr{\de \cT_\rho}{\de \cT_{\rho'}}
    (\bx)
    =
    \prod_{i=1}^n
    \fr{x_i^\dagger \rho x_i}{x_i^\dagger \rho' x_i}
\]
is well defined, since $\mu$ is supported on full-rank matrices.
Let $\nu_{\bx}$ denote the posterior distribution of $\rho$ given observations $\bx$.
The density ratio of any $\rho,\rho' \in S_{\supp}$ under $\nu_{\bx}$ is given by Bayes' rule, and equals
\[
    \fr{\nu_{\bx}(\rho)}{\nu_{\bx}(\rho')}
    = \fr{\de \cT_\rho}{\de \cT_{\rho'}} (\bx)
    \cdot \fr{\mu(\rho)}{\mu(\rho')}.
\]
So, for an arbitrary reference state $\rho' \in S_{\supp}$ (below we take $\rho' = \rho_0$, the unknown true state) the density of $\nu_{\bx}$ is
\[
    \nu_{\bx}(\rho)
    =
    \fr{1}{Z_{\bx}}
    \fr{\de \cT_\rho}{\de \cT_{\rho'}} (\bx)
    \mu(\rho),
    \qquad
    Z_{\bx}
    =
    \int_{U}
    \fr{\de \cT_\rho}{\de \cT_{\rho'}} (\bx)
    \mu(\rho)
    ~\de \Leb_U(\rho).
\]
The main technical component of the proof is the following anti-concentration result for $\nu_\bx$.

\begin{definition}
    For $\rho \in U$, let $B(\rho, \eps)$ denote the ball $\{\rho' \in U : \norm{\rho'-\rho}_{\tr} \le \eps\}$.
    Similarly for $\rho \in U_0$, let $B(\rho, \eps) = \{\rho' \in U_0 : \norm{\rho'-\rho}_{\tr} \le \eps\}$.
\end{definition}
\begin{theorem}
    \label{thm:main}
    Suppose $d \gg 1$, $\eps \le \eps_0$ for an absolute constant $\eps_0$, and $n \ll d^3/\eps^2$.
    If $\rho \in S_{\good}$ and
    % Let $\rho_0 \sim \mu$; on the event $\rho_0 \in S_{\good}$ the following holds.
    % If
    $\bx \sim \cT_{\rho_0}$, there is an event $S_{\rho_0} \in \sigma(\bx)$, with $\bbP(\bx \in S_{\rho_0}) \ge 1-\exp(-d^2)$, on which $\nu_\bx(B(\rho_0,\eps)) \ll 1$.
\end{theorem}
% \begin{proof}
% Note that for all $\rho \in \supp(\mu)$,
% \[
%     0\le
%     \fr{d^3}{4\sigma^2} \norm{\rho-\fr{1}{d}I}_F^2
%     \le
%     \fr{d^4}{4\sigma^2}\norm{\rho-\fr{1}{d}I}_{\op}^2
%     \le
%     4d^2,
% \]
% so for any $\rho,\rho' \in \supp(\mu)$, $\mu(\rho) / \mu(\rho') \ge \exp(-4d^2)$.

Let $C$ be a large constant we will set later and $\eps_0 = \sigma / C$.
The starting point of the proof of Theorem~\ref{thm:main} is the estimate
\begin{align}
    \label{eq:likelihood-ratio-0}
    \fr{\nu_\bx(B(\rho_0,C\eps))}{\nu_\bx(B(\rho_0,\eps))}
    &=
    \fr{
        \int_{B(\rho_0, C\eps)}
        \fr{\de \cT_\rho}{\de \cT_{\rho_0}}(\bx)
        \mu(\rho)
        ~\de \Leb_U(\rho)
    }{
        \int_{B(\rho_0,\eps)}
        \fr{\de \cT_\rho}{\de \cT_{\rho_0}}(\bx)
        \mu(\rho)
        ~\de \Leb_U(\rho)
    } \\
    &\ge
    \label{eq:likelihood-ratio-1}
    \exp(-4d^2)
    \fr{
        \int_{B(\rho_0, C\eps)}
        \fr{\de \cT_\rho}{\de \cT_{\rho_0}}(\bx)
        \ind\{\rho \in S_{\supp}\}
        ~\de \Leb_U(\rho)
    }{
        \int_{B(\rho_0,\eps)}
        \fr{\de \cT_\rho}{\de \cT_{\rho_0}}(\bx)
        \ind\{\rho \in S_{\supp}\}
        ~\de \Leb_U(\rho)
    },
\end{align}
where the second line uses Lemma~\ref{lem:mu-ratio}.
Applying Lemma~\ref{lem:mu-ratio} in this way amounts to replacing $\mu$ in the numerator of \eqref{eq:likelihood-ratio-0} with a measure that sub-samples it, and in the denominator with a measure that upper bounds it.
Define the volumes
\[
    V_1 = \int_{B(0,1)}~\de \Leb_{U_0}(\rho),
    \qquad
    V_2 = \int_{B(0,1)} \ind\lt\{\norm{\rho}_{\op} \le 1/d\rt\} ~\de \Leb_{U_0}(\rho).
\]
We now separately bound the numerator and denominator of \eqref{eq:likelihood-ratio-1} in terms of these volumes, beginning with the denominator.
\begin{lemma}
    \label{lem:denominator}
    If $\rho_{0} \in S_{\good}$, there is an event $S_{\rho_0} \in \sigma(\bx)$, with $\Pr(\bx \in S_{\rho_0}) \ge 1 - \exp(-d^2)$, on which
    \[
        \int_{B(\rho_0,\eps)}
        \fr{\de \cT_\rho}{\de \cT_{\rho_0}}(\bx)
        \ind\{\rho \in S_{\supp}\}
        ~\de \Leb_U(\rho)
        \le \exp(d^2) \eps^{(d+2)(d-1)/2} V_1\,.
    \]
\end{lemma}
\begin{proof}
    Note that $\E_{\bx \sim \cT_{\rho_0}} \fr{\de T_\rho}{\de T_{\rho_0}}(\bx) = 1$.
    So
    \begin{align*}
        \E_{\bx \sim \cT_{\rho_0}}
        \int_{B(\rho_0,\eps)}
        \fr{\de \cT_\rho}{\de \cT_{\rho_0}}(\bx)
        \ind\{\rho \in S_{\supp}\}
        ~\de \Leb_U(\rho)
        &=
        \int_{B(\rho_0,\eps)}
        \ind\{\rho \in S_{\supp}\}
        ~\de \Leb_U(\rho) \\
        &\le
        \int_{B(\rho_0,\eps)}
        ~\de \Leb_U(\rho)
        =
        \eps^{(d+2)(d-1)/2} V_1.
    \end{align*}
    The exponent $(d+2)(d-1)/2$ comes from the fact that the space of symmetric matrices has dimension $d(d+1)/2$, so $U$ has dimension $d(d+1)/2 - 1 = (d+2)(d-1)/2$.
    The result follows from Markov's inequality.
\end{proof}
Before bounding the numerator of \eqref{eq:likelihood-ratio-1}, we define the set
\[
    N_\ast(\rho_0) = \lt\{
        \rho \in U :
        \norm{\rho - \rho_0}_{\op} \le \fr{C\eps}{d},
        \norm{\rho - \rho_0}_{\tr} \le C\eps
    \rt\},
\]
which is an isotropic neighborhood of $\rho_0$.
Let $\gamma(\rho_0)$ denote the uniform distribution on $N_\ast(\rho_0)$ (w.r.t. $\Leb_U$).
That is, for bounded measurable test function $f : U \to \bbR$,
\[
    \E_{\rho \sim \gamma(\rho_0)} f(\rho)
    =
    \fr{
      \int_U f(\rho)
      \ind\{\rho \in N_\ast(\rho_0)\}
      \de \Leb_U(\rho)
    }{
      \int_U
      \ind\{\rho \in N_\ast(\rho_0)\}
      \de \Leb_U(\rho)
    }.
\]
\begin{lemma}
    \label{lem:numerator}
    If $\rho_0 \in S_{\good}$ and $\eps \le \eps_0$, then
    \begin{equation}
      \label{eq:numerator-bd}
      \int_{B(\rho_0, C\eps)}
      \fr{\de \cT_\rho}{\de \cT_{\rho_0}}(\bx)
      \ind\{\rho \in S_{\supp}\}
      ~\de \Leb_U(\rho)
      \ge
      (C\eps)^{(d+2)(d-1)/2}
      \E_{\rho \sim \gamma(\rho_0)}\lt[\fr{\de \cT_\rho}{\de \cT_{\rho_0}}(\bx)\rt]
      V_2.
    \end{equation}
\end{lemma}
\begin{proof}
    Because $\rho_0 \in S_{\good}$, we have $\norm{\rho_0-\fr1d I_d}_\op \le \fr{3\sigma}{d}$.
    Because $\eps \le \eps_0 = \sigma/C$, if $\rho$ satisfies $\norm{\rho-\rho_0}_\op \le \fr{C\eps}{d}$ we have the implication chain
    \[
      \norm{\rho-\rho_0}_\op \le \fr{C\eps}{d}
      \Rightarrow
      \norm{\rho-\rho_0}_\op \le \fr{\sigma}{d}
      \Rightarrow
      \norm{\rho-\fr1d I_d}_\op \le \fr{4\sigma}{d}
      \Rightarrow
      \rho \in S_{\supp}.
    \]
    So, letting $X$ denote the left-hand side of \eqref{eq:numerator-bd}, we have
    \begin{align*}
      X &\ge
      \int_{B(\rho_0, C\eps)}
      \fr{\de \cT_\rho}{\de \cT_{\rho_0}}(\bx)
      \ind\lt\{\norm{\rho-\rho_0}_\op \le \fr{C\eps}{d}\rt\}
      ~\de \Leb_U(\rho)
      =
      \int_{U}
      \fr{\de \cT_\rho}{\de \cT_{\rho_0}}(\bx)
      \ind\{\rho \in N_\ast(\rho_0)\}
      ~\de \Leb_U(\rho) \\
      &=
      \E_{\rho \sim \gamma(\rho_0)}
      \lt[\fr{\de \cT_\rho}{\de \cT_{\rho_0}}(\bx)\rt]
      \int_U \ind\{\rho \in N_\ast(\rho_0)\}
      ~\de \Leb_U(\rho).
    \end{align*}
    Finally, note that
    \begin{align*}
      \int_U \ind\{\rho \in N_\ast(\rho_0)\}
      ~\de \Leb_U(\rho)
      =
      \int_{B(\rho_0, C\eps)}
      \ind\lt\{\norm{\rho - \rho_0}_{\op} \le \fr{C\eps}{d}\rt\}
      ~\de \Leb_U(\rho)
      = (C\eps)^{(d+2)(d-1)/2} V_2,
    \end{align*}
    which conludes the proof.
\end{proof}
It remains to control $\E_{\rho \sim \gamma(\rho_0)} \lt[\fr{\de \cT_\rho}{\de \cT_{\rho_0}}(\bx)\rt]$ and the volume ratio $V_2/V_1$.
The former measures the information gained from the observations $\bx$.
It is bounded by the following lemma, which formalizes the intuition that each observation improves the log likelihood ratio by at most $O(\eps^2/d)$.
This is the step that uses the hypothesis $n \ll d^3/\eps^2$.
\begin{lemma}
    \label{lem:likelihood-ratio-improvement}
    If $\rho_0 \in S_{\good}$, then for any sequence of unit vectors $\bx = (x_1,\ldots,x_n)$,
    % Under the assumptions of Theorem~\ref{thm:main},
    \[
        \E_{\rho \sim \gamma(\rho_0)}
        \lt[\fr{\de \cT_\rho}{\de \cT_{\rho_0}}(\bx)\rt]
        \ge
        \exp(-d^2).
    \]
\end{lemma}
\begin{proof}
    For all $\rho\in S_{\supp}$, the eigenvalues of $\rho$ lie within $[0.96/d, 1.04/d]$.
    Thus, for any unit vector $x$, $\fr{x^\dagger \rho x}{x^\dagger \rho_0 x} \in [0.96/1.04, 1.04/0.96] \subseteq [0.9, 1.1]$.
    Using the fact that $\log(1+a) \ge a - \fr23 a^2$ for $|a|\le 0.1$, we have
    \[
        \log \fr{x^\dagger \rho x}{x^\dagger \rho_0 x}
        \ge
        \fr{x^\dagger (\rho-\rho_0) x}{x^\dagger \rho_0 x}
        - \fr23
        \lt(\fr{x^\dagger (\rho-\rho_0) x}{x^\dagger \rho_0 x}\rt)^2
        \ge
        \fr{x^\dagger (\rho-\rho_0) x}{x^\dagger \rho_0 x}
        -d^2 (x^\dagger (\rho - \rho_0) x)^2.
    \]
    By symmetry, $\E_{\rho \sim \gamma(\rho_0)} (\rho - \rho_0) = 0$, and by rotational invariance,
    \[
        d^2\E_{\rho \sim \gamma(\rho_0)}
        \lt[(x^\dagger (\rho - \rho_0) x)^2 \rt]
        =
        \E_{\rho \sim \gamma(\rho_0)}
        \lt[\norm{\rho - \rho_0}_F^2\rt]
        \le
        \E_{\rho \sim \gamma(\rho_0)}
        \lt[
            \norm{\rho - \rho_0}_{\tr}
            \norm{\rho - \rho_0}_{\op}
        \rt]
        \le
        \fr{C^2\eps^2}{d}.
    \]
    By Jensen's inequality and the above estimates,
    \begin{align*}
        \log \E_{\rho \sim \gamma(\rho_0)} \lt[\fr{\de \cT_\rho}{\de \cT_{\rho_0}}(\bx)\rt]
        &\ge
        \sum_{i=1}^n
        \E_{\rho \sim \gamma(\rho_0)}
        \log \fr{x_i^\dagger \rho x_i}{x_i^\dagger \rho_0 x_i} \\
        &\ge
        \sum_{i=1}^n
        \E_{\rho \sim \gamma(\rho_0)} \lt[
            \fr{x_i^\dagger (\rho - \rho_0)x_i}{x_i^\dagger \rho_0 x_i}
            - d^2 (x_i^\dagger (\rho - \rho_0) x_i)^2
        \rt] \\
        &\ge
        -\fr{C^2n\eps^2}{d}.
    \end{align*}
    Finally, because $n \ll d^3/\eps^2$, this is lower bounded by $-d^2$.
\end{proof}
The volume ratio $V_2/V_1$ is bounded by the following lemma, whose proof we defer to Section~\ref{sec:volume-ratio}.
The proof uses tools from random matrix theory.
\begin{lemma}
    \label{lem:volume-ratio}
    We have that $V_2/V_1 \ge \exp(-3d^2)$.
\end{lemma}
We now put the above claims together to prove Theorem~\ref{thm:main}.
\begin{proof}[Proof of Theorem~\ref{thm:main}]
    Let $S_{\rho_0}$ be the event from Lemma~\ref{lem:denominator}.
    By the calculation \eqref{eq:likelihood-ratio-1} and Lemmas~\ref{lem:denominator} and \ref{lem:numerator}, for all $\bx \in S_{\rho_0}$,
    \[
        \fr{\nu_\bx(B(\rho_0,C\eps))}{\nu_\bx(B(\rho_0,\eps))}
        \ge
        \exp(-5d^2)
        C^{(d+2)(d-1)/2}
        \E_{\rho \sim \gamma(\rho_0)}
        \lt[\fr{\de \cT_\rho}{\de \cT_{\rho_0}}(\bx)\rt]
        \cdot \fr{V_2}{V_1}.
    \]
    Lemmas~\ref{lem:likelihood-ratio-improvement} and \ref{lem:volume-ratio} bound the remaining factors, giving
    \[
        \fr{\nu_\bx(B(\rho_0,C\eps))}{\nu_\bx(B(\rho_0,\eps))}
        \ge
        \exp(-9d^2)
        C^{(d+2)(d-1)/2}.
    \]
    Taking $C = e^{20}$ gives $\fr{\nu_\bx(B(\rho_0,C\eps))}{\nu_\bx(B(\rho_0,\eps))} \gg 1$.
    Since $\nu_\bx(B(\rho_0,C\eps)) \le 1$, this implies $\nu_\bx(B(\rho_0,\eps)) \ll 1$.
\end{proof}

\subsection{Main Lower Bound}

We can now prove our main lower bound for tomography with incoherent measurements, which we state formally below:

\begin{theorem}\label{thm:tomo_lbd}
    There exist absolute constants $\eps_0 > 0$ and $d_0\in\mathbb{N}$ such that for any $0 < \eps < \eps_0$ and any integer $d\ge d_0$, the following holds. If $n = o(d^3/\eps^2)$, then for any algorithm for state tomography  $(\cT,\cA)$ that uses $n$ incoherent, possibly adaptive, measurements, its output $\widehat{\rho}$ upon measuring $n$ copies of $\rho$ satisfies $\norm{\rho - \widehat{\rho}}_{\tr} > \eps$ with probability $1 - o(1)$.
\end{theorem}

\begin{proof}
    Let $S\in\sigma(\rho,\bx)$ be the event that $\rho\sim\mu$ lies in $S_{\good}$ and $\bx\sim \cT_{\rho}$ lies in $S_{\rho}$. In this proof we will abuse notation and use $\cA$ to also denote the internal randomness used by $\cA$. It suffices to show $\Pr_{\cA,\rho\sim\mu,\bx\sim\cT_\rho}[\norm{\cA(\bx) - \rho}_{\tr} \leq \eps] = o(1)$.

    First note that
    \begin{align}
        \Pr_{\cA,\rho,\bx}\left[\norm{\cA(\bx) - \rho}_{\tr} \le \eps \right] &= \E_{\cA,\bx}\E_{\rho\sim\nu_{\bx}}\left[\bone{\norm{\cA(\bx) - \rho}_{\tr} \le \eps}\right] \\
        &\le \E_{\cA,\bx}\E_{\rho\sim\nu_{\bx}}\left[\bone{\norm{\cA(\bx) - \rho}_{\tr} \le \eps \ \text{and} \ (\rho,\bx)\in S}\right] + o(1) \label{eq:union}
    \end{align}
    where the second step follows by a union bound and the fact that $\Pr[(\rho,\bx)\not\in S] = e^{-\Omega(d)} + e^{-\Omega(d^2)} = o(1)$ by Theorem~\ref{thm:main}.

    For any choice of internal randomness for $\cA$ and any transcript $\bx$, let $\rho^\cA_\bx$ denote an arbitrary state for which $(\rho^\cA_\bx,\bx)\in S$ and $\norm{\cA(\bx) - \rho^\cA_\bx}_{\tr} \le \eps$, if such a state exists. Denote by $\cE$ the event that such a state exists. Then under $\cE$, for any state $\rho$ for which $\norm{\cA(\bx) - \rho}_\tr \le \eps$, we have $\norm{\rho^\cA_\bx - \rho}_\tr \le 2\eps$. If $\cE$ does not occur for some choice of internal randomness for $\cA$ and some $\bx$, note that the corresponding inner expectation in \eqref{eq:union} is zero.
    We can thus upper bound the double expectation in \eqref{eq:union} by
    \begin{align}
        \E_{\cA,\bx | \cE}\E_{\rho\sim\nu_{\bx}}\left[\bone{\norm{\rho^\cA_\bx-\rho}_{\tr} \le 2\eps \ \text{and} \ (\rho,\bx)\in S}\right] \le \E_{\cA,\bx | \cE}\Pr_{\rho\sim\nu_{\bx}}\left[\norm{\rho^\cA_\bx-\rho}_{\tr} \le 2\eps\right] = o(1),
    \end{align}
    where in the last step we used the fact that under $\cE$ we have $(\rho^\cA_\bx,\bx)\in S$, so by Theorem~\ref{thm:main} the posterior measure $\nu_\bx$ places $o(1)$ mass on the trace norm $\eps$-ball around $\rho^\cA_\bx$.
\end{proof}

\section{Lower Bounding the Volume Ratio: Proof of Lemma~\ref{lem:volume-ratio}}\label{sec:volume-ratio}

In this section, we will prove Lemma~\ref{lem:volume-ratio}, which lower bounds the volume ratio $V_2/V_1$.
\\\\
Let
\[
    V = \lt\{
        \blam = (\lambda_1,\ldots,\lambda_d) \in \bbR^d :
        \sum_{i=1}^d \lambda_i = 0
    \rt\}.
\]
This is a subspace of $\bbR^d$ of codimension $1$.
It inherits the inner product of $\bbR^d$, which defines a Lebesgue measure $\Leb_V$.
Define
\[
    \Delta = \lt\{
        \blam \in V :
        \lambda_1 \ge \lambda_2 \ge \cdots \ge \lambda_d,
        \sum_{i=1}^d |\lambda_i| \le 1
    \rt\},
    \qquad
    \Delta' = \lt\{
        \blam \in \Delta : \max_{i\in [d]} |\lambda_i| \le 1/d
    \rt\},
\]
and
\[
    \Gamma = \lt\{
        \blam \in V :
        \lt|\lambda_i - \fr{d-2i+1}{d^2}\rt|
        \le
        \fr{1}{d^4}
    \rt\}.
\]
\begin{lemma}
    We have $\Gamma \subseteq \Delta'$.
\end{lemma}
\begin{proof}
    If $\blam \in \Gamma$, then
    \[
        \lambda_i - \lambda_{i+1} \ge \fr{2}{d^2} - \fr{2}{d^4} > 0
    \]
    and
    \[
        \sum_{i=1}^d |\lambda_i| \le d\cdot \fr{1}{d} = 1,
    \]
    so $\blam \in \Delta$.
    Moreover
    \[
        |\lambda_i| \le \fr{d-1}{d^2} + \fr{1}{d^4} \le \fr{1}{d}. \qedhere
    \]
\end{proof}
The volume ratio $V_2/V_1$ is the probability that if $\rho$ is drawn from the uniform measure on $B(0,1)$ (w.r.t. $\Leb_{U_0}$), then $\norm{\rho}_\op \le 1/d$.
The main random matrix theory fact we will use is \cite[Theorem 2.5.2]{guionnet2005rmt}, which implies that if $\blam = (\lambda_1,\ldots,\lambda_d)$ are the eigenvalues of $\rho$ drawn from this distribution, then $\blam$ has density (w.r.t. $\Leb_V$) $\fr{1}{Z} f(\blam)$, where $Z$ is a normalizing constant and
\[
    f(\blam)
    =
    \ind\{\blam \in \Delta\}
    \prod_{1\le i<j\le d}
    |\lambda_i - \lambda_j|.
\]
For measurable $A \subseteq V$, let $\Vol(A) = \int_A 1 \de \Leb_V$ denote the volume of $A$.
Then,
\begin{equation}
    \label{eq:volume-ratio-1}
    \fr{V_2}{V_1}
    =
    \fr{\int_{\Delta'} f(\blam) ~\de \Leb_V(\blam)}{\int_\Delta f(\blam) ~\de \Leb_V(\blam)}
    \ge
    \fr{\int_{\Gamma} f(\blam) ~\de \Leb_V(\blam)}{\int_\Delta f(\blam) ~\de \Leb_V(\blam)}
    \ge
    \fr{\Vol(\Gamma)}{\Vol(\Delta)} \cdot
    \fr{\inf_{\blam \in \Gamma} f(\blam)}{\sup_{\blam \in \Delta} f(\blam)}.
\end{equation}
The following three propositions bound the quantities in the right-hand side of \eqref{eq:volume-ratio-1}.
\begin{proposition}
    \label{prop:gamma-f-lb}
    For any $\blam \in \Gamma$, $f(\blam) \ge 1/((2e)^{d^2/2}d^{d(d-1)/2})$.
\end{proposition}
\begin{proof}
    For each $i\in [d]$,
    \begin{align*}
        \prod_{j\in [d]\setminus \{i\}}
        |\lambda_i-\lambda_j|
        &\ge
        \prod_{j=1}^{i-1}
        \lt(\fr{2(i-j)}{d^2} - \fr{2}{d^4}\rt)
        \prod_{j=i+1}^{d}
        \lt(\fr{2(j-i)}{d^2} - \fr{2}{d^4}\rt)
        \ge
        \prod_{j=1}^{i-1}
        \fr{i-j}{d^2}
        \prod_{j=i+1}^{d}
        \fr{j-i}{d^2}
        \ge
        \prod_{j=1}^{d-1}
        \fr{j}{2d^2} \\
        &\ge \fr{(d-1)!}{2^d d^{2(d-1)}}
        =
        \fr{d!}{2^d d^{2d-1}}
        \ge
        \fr{(d/e)^d}{2^d d^{2d-1}}
        \ge
        \fr{1}{(2e)^d d^{d-1}}.
    \end{align*}
    Thus
    \begin{align}
        f(\blam)
        &=
        \lt(\prod_{i=1}^d \prod_{j\in [d]\setminus \{i\}} |\lambda_i-\lambda_j|\rt)^{1/2}
        \ge
        \fr{1}{(2e)^{d^2/2}d^{d(d-1)/2}}. \qedhere
    \end{align}
\end{proof}
\begin{proposition}
    \label{prop:delta-f-lb}
    For any $\blam \in \Delta$, $f(\blam) \le e^{2d^2}/d^{d(d-1)/2}$.
\end{proposition}
\begin{proof}
    Let $(\olam_1, \ldots, \olam_d)$ be the permutation of $(\lambda_1,\ldots,\lambda_d)$ with $|\olam_1| \le \cdots \le |\olam_d|$.
    For each $i\in [d]$,
    \[
        \prod_{j<i}
        |\olam_i-\olam_j|
        \le
        (2|\olam_i|)^{i-1}
        \le
        \lt(\fr{e^{2d|\olam_i|}}{d}\rt)^{i-1}
        \le
        \fr{e^{2d^2|\olam_i|}}{d^{i-1}},
    \]
    so (since $\sum_{i=1}^d |\lambda_i|\le 1$)
    \begin{align}
        f(\blam)
        &=
        \prod_{i=1}^d
        \prod_{j<i}
        |\olam_i-\olam_j|
        \le
        \fr{e^{2d^2 \sum_{i=1}^d |\olam_i|}}{d^{d(d-1)/2}}
        \le
        \fr{e^{2d^2}}{d^{d(d-1)/2}}. \qedhere
    \end{align}
\end{proof}
\begin{proposition}
    \label{prop:delta-gamma-volumes}
    We have $\Vol(\Delta) \le e^{O(d)}$ and $\Vol(\Gamma) \ge d^{-O(d)}$.
\end{proposition}
\begin{proof}
    The projection of $V$ onto its first $d-1$ coordinates has Jacobian $\Theta(1)$.
    Because this projection maps $\Delta$ injectively to $\bbR^{d-1}$ and each resulting coordinate is in $[-1, 1]$,
    \[
        \Vol(\Delta) \le \Theta(1) \cdot 2^{d-1} \le e^{O(d)},
    \]
    proving the first conclusion.
    Note that $\Gamma$ is the set of $\blam$ satisfying $\lambda_i = \fr{d-2i+1}{d^2} + \fr{1}{d^4} \mu_i$, where $\bmu = (\mu_1,\ldots,\mu_d)$ ranges over
    \[
        \Gamma' = \lt\{\bmu \in V : \max_{i\in [d]} |\mu_i| \le 1\rt\}.
    \]
    The set $\Gamma'$ certainly contains the set of $\bmu$ where $|\mu_1|,\ldots,|\mu_{d-1}| \le 1/d$ and $\mu_d = -\sum_{i=1}^{d-1} \mu_i$.
    So,
    \[
        \Vol(\Gamma') \ge \Theta(1) (2/d)^{d-1} = d^{-O(d)}.
    \]
    Finally,
    \[
        \Vol(\Gamma) = \Vol(\Gamma') \cdot (d^{-4})^{d-1} = d^{-O(d)},
    \]
    proving the second conclusion.
\end{proof}
\begin{proof}[Proof of Lemma~\ref{lem:volume-ratio}]
    By equation \eqref{eq:volume-ratio-1} and the last three propositions,
    \[
        \fr{V_2}{V_1}
        \ge
        \fr{d^{-O(d)}}{e^{O(d)}}
        \cdot \fr{(2e)^{-d^2/2}}{e^{2d^2}}
        \ge e^{-3d^2}. \qedhere
    \]
    % The set $\Delta$ has volume w.r.t. $\Leb_V$ at most $e^{O(d)}$, because the projection of $\Delta$ onto its first $d-1$ coordinates is a bijection to $\bbR^{d-1}$ with Jacobian $O(1)$, and each resulting coordinate is in $[-1, 1]$.
    % Similarly, the set $\Gamma$ has volume w.r.t. $\Leb_V$ at least $O(1)\cdot (2d^{-4})^{d-1} = d^{-O(d)}$.
    % By the last two propositions,
    % \[
    %     \fr{V_2}{V_1}
    %     \ge
    %     \fr{\int_\Gamma f(\blam) ~\de \Leb_V(\blam)}{\int_\Delta f(\blam) ~\de \Leb_V(\blam)}
    %     \ge
    %     \fr{d^{-O(d)}}{e^{O(d)}}
    %     \cdot
    %     \fr{\inf_{\blam \in \Gamma}f(\blam)}{\sup_{\blam \in \Delta}f(\blam)}
    %     \ge
    %     \fr{d^{-O(d)}}{e^{O(d)}}
    %     2^{-d^2/2}
    %     e^{-5d^2/2}
    %     \ge
    %     e^{-3d^2}. \qedhere
    % \]
\end{proof}

\iffalse
\documentclass{article}
\usepackage[utf8]{inputenc}
\usepackage{preamble}
\usepackage{enumitem}
\usepackage{fullpage}
\newcommand*{\defeq}{\stackrel{\text{def}}{=}}

\newcommand{\rhat}{\widehat{\rho}}
\newcommand{\tr}{\mathrm{tr}}
\newcommand{\op}{\mathrm{op}}
\newcommand{\diag}{\mathrm{diag}}
\newcommand{\Gbar}{\overline{\Gamma}}
\newcommand{\sitan}[1]{{\color{red} [sitan: #1]}}

\title{Tomography in Infidelity}
\author{}
\date{January 2023}

\begin{document}

\maketitle

\begin{enumerate}
    \item Nonadaptive algorithm achieves $\sqrt{d/n}$ operator norm closeness (for entangled is the scaling $\sqrt{1/n}$???)
    \item Series expansion: bounds infidelity between matrix and matrix without off-diagonal blocks (jamboard P. 17)
    \item Triangle inequality in infidelity
    \item Recursive partitioning: take crude estimator and partition into eigenspaces above and below $1/d$, then recurse and partition into eigenspaces above and below $1/2d$, etc., all the way down to $\gamma/d$ (mix with $\gamma/d \cdot I$ to ensure true state doesn't have smaller eigenvalues)
\end{enumerate}
\fi

\section{Basic Learning Results}
% \allen{Appendix this section?}

Recall that for two quantum states $\rho, \sigma$, the fidelity between them is $F(\rho, \sigma) = \tr (\sqrt{\sqrt{\rho} \sigma \sqrt{\rho}})^2$.  This is the quantum analogue of the \emph{Bhattacharyya coefficient}, which for two classical probability distributions $p, q$ over $[d]$, is defined to be $BC(p, q) = \sum_{i = 1}^d \sqrt{p_i} \sqrt{q_i}$.
Note that if $\rho$ and $\sigma$ commute, then $F(\rho, \sigma) = BC(p, q)^2$, where $p, q$ are the classical distributions given by the eigenvalues of $\rho$ and $\sigma,$ respectively.
The following inequality is well known:

\begin{fact}
\[
BC (p, q) \geq 1 - O \left( \chi^2 (p\| q) \right) \; .
\]

\end{fact}

As an immediate corollary of this, we have the following:
\begin{corollary}
\label{cor:mix}
Let $\rho \in \co^{d \times d}$ be an arbitrary mixed state, and let $\tilde{\rho} = (1 - \gamma) \rho + \gamma \cdot \tfrac{1}{d} \Id$.
Then $F(\rho, \tilde{\rho}) \geq 1 - O \left( \gamma \right)$.
\end{corollary}
\begin{proof}
Let $p$ and $\tilde{p}$ denote the the distributions given by the eigenvalues of $\rho$ and $\tilde{\rho}$, respectively.
Since $\rho$ and $\tilde{\rho}$ commute, it suffices to show that $BC (p, \tilde{p}) \geq 1 - O(\gamma)$.
However, we have:
\begin{align*}
    \chi^2 (\tilde{p}\| p) &= \sum_{i = 1}^d \frac{(\gamma p_i + \gamma / d)^2}{\tilde{p}_i} \leq 2 \sum_{i = 1}^d \frac{\gamma^2 p_i^2}{\tilde{p_i}} + 2 \sum_{i = 1}^d \frac{\gamma^2}{d^2 \tilde{p}_i} \leq O(\gamma) \; ,
\end{align*}
since $\tilde{p}_i \geq \gamma / d$ for all $i$.
\end{proof}

It is well-known that infidelity is not a metric on mixed states.
However, the associated Bures metric, defined as
\[
    D_B (\rho, \sigma)^2 = 2 \left( 1 - \sqrt{F(\rho, \sigma)} \right) \; ,
\]
does satisfy the triangle inequality and is hence a valid metric.
As an immediate consequence of this, we obtain that infidelity still satisfies a weak version of the triangle inequality:
\begin{corollary}
\label{cor:fidelity-triangle}
    Let $k$ be a positive integer, and let $\gamma \ll 1/ k^2$ be sufficiently small.
    Let $\rho_1, \ldots, \rho_k$ be a sequence of mixed states satisfying $1 - F(\rho_i, \rho_{i + 1}) \leq \gamma$ for all $i = 1, \ldots, k - 1$.
    Then 
    \[
        1 - F(\rho_1, \rho_k) \leq O(k^2 \gamma) \; .
    \]
\end{corollary}
\begin{proof}
    By repeated application of the Taylor series expansion of the square root function around $1$, we know that $D_B (\rho_i, \rho_{i + 1}) \leq O(\sqrt{\gamma})$ for all $i = 1, \ldots, k - 1$.
    Therefore by the triangle inequality, we have that $D_B (\rho_1, \rho_k) \leq O(k \sqrt{\gamma})$, from which the claim immediately follows.
\end{proof}

\subsection{Learning a state in spectral norm}
Finally, we require the following guarantee for one of the standard estimators for a mixed state based on unentangled measurements.
Given $n$ copies of a mixed state $\rho$, we measure each one with the uniform POVM over the sphere $\{d \ket{v} \bra{v} dv \}$ where $v$ ranges over the unit sphere, and let $\ket{v_i}$ denote the outcome of the $i$th measurement.
We consider the estimator $H_n (\rho) = H_n (\rho, v_1, \ldots, v_n)$, defined as
\begin{equation}
\label{eq:spectral-estimator}
    H_n (\rho) = \frac{1}{n} \sum_{i = 1}^n \left( (d + 1) \ket{v_i} \bra{v_i} - \Id  \right) \; .
\end{equation}
We show the following rate for this estimator. The same rate is claimed in the proof sketch of Theorem 2 in~\cite{guctua2020fast}, but to our knowledge no full proof of this exists in the literature. We include a full proof for completeness:
\begin{theorem}
\label{thm:spectral-estimator}
There exists a universal constant $C$ so that for all $n$, we have that
\[
    {\norm{H_n (\rho) - \rho}_{\op}} \leq C \cdot \max \left( \frac{d + \log 1 / \delta}{n}, \sqrt{\frac{d + \log 1 / \delta}{n}} \right)  \; ,
\]
with probability $1 - \delta$.
\end{theorem}
\noindent
The key concentration lemma we require is the following:
\begin{lemma}
\label{lem:subexp-concentration}
Let $\ket{v}$ be the outcome of measuring $\rho$ using the uniform POVM.
Then, for any fixed pure state $\ket{u}$, and for all $k \geq 1$, we have
\[
    \E \brk*{(d + 1)^k \braket{u | v}^{2k} } \leq (k + 1)^{k + 1} \; .
\]
\end{lemma}
\begin{proof}
    First consider the case where $\rho = \ket{w} \bra{w}$ is a pure state.
    For any $t$, let $\Pi_t$ denote projection onto the $t$-fold symmetric subspace.
    Then, we have:
    \begin{align}
        \E \brk*{\braket{u | v}^{2k} } &= d \cdot \int \braket{u | v}^{2k} \braket{v | w}^2 dv \\
        &= d \cdot \bra{u}^{\otimes k} \otimes \bra{w} \left( \int \ket{v}^{\otimes (k + 1)} \bra{v}^{\otimes (k + 1)} dv \right) \ket{u}^{\otimes k} \otimes \ket{w} \\
        &= d \cdot \binom{k + d}{k + 1}^{-1} \cdot \left( \bra{u}^{\otimes k} \otimes \bra{w} \right) \Pi_{k + 1} \left( \ket{u}^{\otimes k} \otimes \ket{w} \right) \leq d \cdot \binom{k + d}{k + 1}^{-1} \; ,
\end{align}
where the third step follows by the standard Haar integral formulation of $\Pi_t$ \cite{harrow2013church}.
Therefore, we have that
\begin{align}
    \E \brk*{(d + 1)^k \braket{u | v}^{2k} } &\leq (d + 1)^{k + 1} \cdot \binom{k + d}{k + 1}^{-1} \leq (d + 1)^{k + 1} \cdot \frac{(k + 1)^{k + 1}}{(d + k)^{k + 1}} \leq (k + 1)^{k + 1} \; ,
\end{align}
as claimed.
The claim for general $\rho$ directly follows by convexity.
\end{proof}
\begin{proof}[Proof of Theorem~\ref{thm:spectral-estimator}]
    The proof proceeds via the same general strategy as in~\cite{guctua2020fast}.
    Let $\calN$ be a $1/3$-net of all pure states in $\co^d$.
    For any $u \in \calN$, Lemma~\ref{lem:subexp-concentration} implies that the random variable 
    \[
        \bra{u} \left(\rho -H_n (\rho) \right) \ket{u} = \frac{1}{n} \sum_{i = 1}^n \left( (d + 1) \braket{u, v_i}^2 - 1 - \bra{u} \rho \ket{u} \right)
    \]
    is a sum of $n$ independent $O(1)$-subexponential random variables.
    Therefore, by standard net arguments, we have that for all $\gamma > 0$,
    \[
        \Pr \brk*{  \norm{\rho - H_n (\rho)}_{\op} > \gamma } \leq \exp \left(c_1 d - c_2 n \max (\gamma, \gamma^2) \right) \; ,
    \]
    for some universal constants $c_1, c_2$, which is equivalent to what we wanted to show.
\end{proof}

\noindent
Finally, we also require the following generalization of the estimator we considered above:
\begin{definition}\label{def:projected-POVM}
Given a projection matrix $\Pi \in \co^{d \times d}$ onto an $r$-dimensonal subspace, the \emph{projected estimator} on the subspace $\Pi$, denoted $H_n (\rho, \Pi)$ is defined as follows.
Let $\calM_\Pi = \{r \ket{v} \bra{v} dv \}$ denote the uniform POVM over the subspace spanned by $\Pi$, and consider the POVM over $\co^{d \times d}$ defined by $\calM = \{I - \Pi\} \cup \calM_\Pi$.
Given $n$ copies of $\rho$, let $\ket{v_i}$ denote the outcomes of measuring each copy of $\rho$ independently with $\calM$, where we say $v_i = \bot$ if the outcome is $I - \Pi$.
Then, we define
\[
    H_n (\rho, \Pi) = \frac{1}{n} \sum_{v_i \neq \bot} \left( (r + 1) \ket{v_i} \bra{v_i} - \Id \right) \; .
\]
\end{definition}
\noindent For projected estimators we have the following rate, which follows immediately from Theorem~\ref{thm:spectral-estimator} and standard Chernoff bounds:
\begin{lemma}
\label{lem:spectral-projected}
    Let $\rho \in \co^{d \times d}$ and let $\Pi$ be a projection onto an $r$-dimensional subspace.
    Let $\alpha = \tr (\Pi \rho \Pi)$.
    Then, there exists a universal constant $C$ so that
    \[
    \norm{H_{n} (\rho, \Pi) - \Pi \rho \Pi}_{\op} \leq  C \cdot \max \left( \frac{d + \log 1 / \delta}{n}, \sqrt{\frac{\alpha (d + \log 1 / \delta)}{n}} \right) \; .
    \]
    with probability $1 - \delta$.
\end{lemma}

\section{Learning a state in fidelity}
% \allen{Define $\lesssim$ notation}

In this section, we present our algorithm for tomography in fidelity.  Our main theorem is stated below.
\begin{theorem}\label{thm:learn-in-fidelity}
Let $\rho \in \co^{d \times d}$ be an unknown rank-$r$ mixed state.  Given $n = \wt{O}(dr^2 \log(1/\delta)/\gamma)$ copies of $\rho$, there is an algorithm that uses incoherent measurements and with probability $1 - \delta$, outputs a state $\rhat$ such that  $F(\rho, \rhat) \geq 1 - \gamma$.
\end{theorem}

Let $\eta > 0$ be a parameter to be determined later, and let $t$ be an integer satisfying $t \leq \log_2 r / \gamma + 4$.  We first describe our algorithm.  Divide the samples into $t$ groups of $n / t$ samples each.

We will compute a sequence of orthogonal projections $\Pi_0 = 0, \Pi_1, \ldots, \Pi_t$.
To find $\Pi_{j}$ given $\Pi_0, \ldots, \Pi_{j - 1}$, the algorithm proceed as follows: it forms $\Gamma_j = \Id - \sum_{i = 0}^{j-1} \Pi_i$, and using a fresh batch of samples, it computes $\sigma_{j} = H_{n / t} (\rho, \Gamma_j)$, and it sets $\Pi_{j}$ to be the span of the nonzero eigenvectors of $\sigma_j$ with corresponding eigenvalue at least $2^{-j}$.
When it finishes, it outputs the state
\begin{equation}
M_n (\rho) \defeq \frac{\widehat{\sigma}}{\tr (\widehat{\sigma})} \; , \; \mbox{where} \; \widehat{\sigma} \defeq \sum_{i = 1}^t \Pi_i \sigma_i \Pi_i \; ,    
\end{equation}
To analyze the algorithm, we make the following definitions.
Let $\Pi_{t + 1} = \Gamma_{t + 1}$, and define the state
\begin{equation}
    \rho_{\diag} = \sum_{i = 1}^{t + 1} \Pi_{i} \rho \Pi_i \; .
\end{equation}
For all $i = 1, \ldots, t + 1$, let $B_i \in \co^{d \times \mathrm{rank}(\Pi_i)}$ be any matrix with orthonormal columns so that $B_i B_i^\top = \Pi_i$.
Similarly, for $i = 1, \ldots, t + 1$, let $C_i \in \co^{d \times \mathrm{rank} (\Gamma_i)}$ be any matrix with orthonormal columns so that $C_i C_i^\top = \Gamma_i$.

Notice that by construction $\sum_{i = 1}^{t + 1} \Pi_i = \Id$, so this is indeed a valid mixed state.
Furthermore, note that $\rho_\diag$ is block-diagonal with respect to the matrices $\Pi_1, \ldots, \Pi_{t + 1}$, that is, after a suitable rotation which sends each $B_i$ to itself, $\rho_\diag$ has the form
\[
\rho_\diag = \begin{pmatrix}
    \rho_1 & & &\\
    & \rho_2 & &\\
    & & \ddots &\\
    & & & \rho_{t + 1} \\
\end{pmatrix} \; ,
\]
where we let $\rho_i$ denote the projection of $\Pi_i \rho \Pi_i$ onto its nonzero eigenvectors.
Note that in this basis, $\rho$ is not block diagonal, and indeed, much of the work in the proof is to bound the contribution of the error because of these off-diagonal blocks.
To this end, for all $i = 1, \ldots, t$, let $E_i = B_i^\top \rho C_{i + 1}$, so that
\[
\rho = \begin{pmatrix}
    & \rho_1 & \cellcolor{green} & \cellcolor{green} E_1 & \cellcolor{green}  & \\
    & \cellcolor{green} & \rho_2 & \cellcolor{cyan} \hspace{10pt} E_2 & \cellcolor{cyan}  & \\
    & \cellcolor{green} E_1^\top & \cellcolor{cyan} E_2^\top & \ddots & & \\
    & \cellcolor{green} & \cellcolor{cyan} & & \rho_{t}  & \\
\end{pmatrix} \,,
\]
i.e., the $E_i$ are the off-diagonal components of $\rho$.

Next, since each $\sigma_j$ is computed with fresh samples, by Lemma~\ref{lem:spectral-projected} and a union bound, we have that
\begin{equation}
\label{eq:conc-assume}
\norm{\sigma_j - \Gamma_j \rho \Gamma_j}_{\op} \leq \gamma_j \defeq C \cdot \max \left( \frac{d + \log t / \delta}{n}, \sqrt{\frac{\alpha_j (d + \log t / \delta)}{n}} \right) \; ,
\end{equation}
for all $j = 1, \ldots, t$ simultaneously, with probability at least $1 - \delta$, where $\alpha_j = \tr(\Pi_j \rho \Pi_j)$.

For the remainder of the section, let us take $n \gtrsim \tfrac{(d + \log t / \delta) r^2}{\gamma}$, and we will assume that~\eqref{eq:conc-assume} holds.
We first show a few basic inequalities.  We have the following estimate on the RHS of~\eqref{eq:conc-assume}.
\begin{lemma}
\label{lem:eps-bound}
 Let $n \gtrsim \tfrac{(d + \log t / \delta) r^2}{\gamma}$.
 Then, we have that $\gamma_j \lesssim 2^{-(j + t) / 2}$. 
 % \sitan{maybe more convenient to say $n \ge C\cdot \tfrac{(d + \log t / \delta) r^2}{\gamma}$ for sufficiently large $C > 0$ so that $\gamma_j \le 2^{-j}$ rather than $\gamma_j \lesssim 2^{-j}$?}
\end{lemma}
\begin{proof}
    By our choice of $n$, we have that:
    \[
     \frac{d + \log t / \delta}{n} \lesssim \frac{\gamma}{r^2} \leq 2^{-(j + t) / 2} \; , 
    \]
    and
    \begin{align*}
   \sqrt{\frac{ \alpha_{j - 1} (d + \log t / \delta)}{n}} &= \sqrt{\tr (\Gamma_{j - 1} \rho \Gamma_{ j - 1}) \cdot  \frac{d + \log t / \delta}{n}} \\
   &\lesssim \sqrt{2^{-(j - 1)} \cdot \frac{\gamma}{r}} \leq 2^{-(j + t) / 2} \; ,
    \end{align*}
    where in both cases we use that $t = \log_2 r / \gamma + 4$.
\end{proof}
\noindent With this, we can show the following set of useful inequalities:
\begin{lemma}
\label{lem:projected-norm}
    For all $j = 1, \ldots, t$, we have that 
    \begin{enumerate}[label=(\roman*)]
        \item $\norm{\Gamma_j \rho \Gamma_j}_\op \leq 2^{-(j - 1)}$,
        \item $\norm{\sigma_j}_\op \leq 2^{-(j - 2)}$, and
        \item $B_i^\top \rho B_i \succeq 2^{-j - 1} \Id$, and
        \item $\tr (\Gamma_{t + 1} \rho \Gamma_{t + 1}) \leq \gamma / 2$.
    \end{enumerate}
\end{lemma}
\begin{proof}
    To prove the first bullet point, we proceed by induction.
    The case where $j = 1$ is trivial.
    Now, suppose the claim holds for some $j - 1$.
    By~\eqref{eq:conc-assume}, we have that 
    \[
    \norm{\sigma_{j - 1} - \Gamma_{j - 1} \rho \Gamma_{j - 1} }_{\op} \leq \gamma_{j - 1} \; .
    \]
    Now $\Gamma_j$ is defined to be the projection onto the eigenvectors of $\sigma_{j - 1}$ with eigenvalue less than $2^{-j}$. 
    % \sitan{should $M$ be $\sigma$?}\allen{Seems like it?}
    Therefore, $\norm{\Gamma_j \rho \Gamma_j}_\op \leq \gamma_{j - 1} + 2^{-j}$.
    The second and third claims then both follow from Lemma~\ref{lem:eps-bound}, and by the definition of $M_j$.
    Finally, the last claim follows because $\rho$ has at most $r$ nonzero eigenvalues, and therefore so does $\tr (\Gamma_{t + 1} \rho \Gamma_{t + 1})$; moreover by the above, each one is at most $2^{-t + 2} \leq \gamma / (2r)$.
\end{proof}
One simple but important consequence of these inequalities is that the subspaces $\Pi_i$ are low dimensional:
\begin{corollary}
For all $i,$ we have $\mathrm{rank} (\Pi_i) \leq r$.
\end{corollary}
\begin{proof}
Note that $\rho$ has $r$ nonzero eigenvalues by assumption, so by Lemma~\ref{lem:projected-norm}, $\sigma_i$ can only have $r$ eigenvalues which are at least $2^{- i - 1}$.
\end{proof}

%\noindent One simple but important consequence of these inequalities is that we can ignore the subspace spanned by $\Gamma_{t + 1}$:
%\begin{corollary}
%    Let 
%    \[
%        \rho' = \frac{(\Id - \Gamma_{t + 1}) \rho (\Id - \Gamma_{t + 1})}{\tr ((\Id - \Gamma_{t + 1}) \rho (\Id - \Gamma_{t + 1}))} \; .
%    \]
%    Then we have that $F(\rho', \rho) \leq \gamma / 2$.
%\end{corollary}
%\begin{proof}
%    We have that 
%    \begin{align*}
%        F(\rho', \rho) = \tr \left( \sqrt{\rho^{1/2} \rho' \rho^{1/2} } \right)^2 &= \frac{1}{\tr ((\Id - \Gamma_{t + 1}) \rho (\Id - \Gamma_{t + 1}))} \cdot \tr \left( \sqrt{\left( (\Id - \Gamma_{t + 1}) \rho (\Id - \Gamma_{t + 1}) \right)^2} \right)^2 \\
%        &= \tr ((\Id - \Gamma_{t + 1}) \rho) \geq 1 - \frac{\gamma}{2} \; .\qedhere
%    \end{align*}
%\end{proof}

\noindent We will show the following two key estimates.  Roughly, Lemma~\ref{lem:on-diag} bounds the contribution to infidelity form the error on the diagonal blocks of $\rho$ and $\rho_\diag$.  Lemma~\ref{lem:off-diag} bounds the contribution to infidelity from the off-diagonal blocks in $\rho$.  Putting them together with the (weak) triangle inequality in Corollary~\ref{cor:fidelity-triangle} will immediately imply Theorem~\ref{thm:learn-in-fidelity}.

\begin{lemma}
\label{lem:on-diag}
    Let $\rho, M_n (\rho)$, and $\rho_\diag$ be as above, and assume that~\eqref{eq:conc-assume} holds. Then, we have that
    \[
        F(M_n (\rho), \rho_\diag) \geq 1 - \gamma t \; 
    \]
\end{lemma}

\begin{lemma}
\label{lem:off-diag}
    Let $\rho, M_n (\rho)$, and $\rho_\diag$ be as above, and assume that~\eqref{eq:conc-assume} holds. Then, we have that
    \[
        F(\rho, \rho_\diag) \geq 1 - \gamma \cdot \min (t, d / r) \; . \; 
    \]
\end{lemma}

\begin{proof}[Proof of Theorem~\ref{thm:learn-in-fidelity}]
Combining Lemma~\ref{lem:on-diag}, Lemma~\ref{lem:off-diag} and Corollary~\ref{cor:fidelity-triangle}, we get
\[
F(M_n(\rho). \rho) \geq 1 - O(\gamma \min(t,d/r))
\]
and since $t \leq \log_2 (r/\gamma) + 4$, we can redefine $\gamma$ appropriately (scaling down by a logarithmic factor) to complete the proof.
\end{proof}

\section{Proof of Lemma~\ref{lem:on-diag}}
We first require the following fact, which is a direct consequence of Proposition 3.1 in~\cite{juditsky2008large}, and direct calculation:
\begin{theorem}
\label{thm:hessian-bound}
Let $f(X): \co^{d \times d}_{\geq 0} \to \R$ be defined by $f(X) = \tr (\sqrt{X})$.
Let $A \in \co^{d \times d}$ be a non-degenerate Hermitian matrix.
Then, for all symmetric matrices $B \in \co^{d \times d}$, we have that
\begin{equation}
\label{eq:hessian-bound}
    \nabla^2 f(A) [B, B] \geq -\frac{1}{4} \tr \left( B A^{-3/2} B \right) 
\end{equation}
where $\nabla^2 f(A)$ denotes the Hessian of $f(A)$ (where we treat the entries of $A$ as variables) and $\nabla^2 f(A) [B, B]$ denotes taking the quadratic form of this Hessian at $B$.
\end{theorem}

\begin{proof}[Proof of Lemma~\ref{lem:on-diag}]
    For conciseness, throughout this proof we will let $M = M_n (\rho)$.
    Let $\Delta = M - \rho_\diag$.
    By Taylor's theorem, we know that there is some $\beta \in [0, 1]$ so that
    \begin{align*}
    F(M_n (\rho), \rho_\diag)  &= \tr \left( \sqrt{M^2 - M^{1/2} \Delta M^{1/2}}\right) \\
    &= 1 + \frac{1}{2} \tr \left( M^{-1} M^{1/2} \Delta M^{1/2} \right) + D^2 f \left( M^2 - \beta M^{1/2} \Delta M^{1/2} \right) \left[ M^{1/2} \Delta M^{1/2}, M^{1/2} \Delta M^{1/2} \right] \\
    &\geq 1 - \frac{1}{4} \tr \left((M^{1/2} \Delta M^{-1/4} ( M + \beta \Delta)^{-3/2} M^{-1/4} \Delta M^{1/2} \right) \; ,
    \end{align*}
    where in the last line we used Theorem~\ref{thm:hessian-bound}.

    Notice that both $M$ and $\Delta$ are block diagonal, and moreover, by construction, they share the same block structure.
    Thus, if we let $M_i = \Pi_i M \Pi_i$ and $\Delta_i = \Pi_i \Delta \Pi_i$, then we have that
    \[
    \tr \left((M^{1/2} \Delta M^{-1/4} ( M + \beta \Delta)^{-3/2} M^{-1/4} \Delta M^{1/2} \right) = \sum_{i = 1}^t \tr \left(M_i^{1/2} \Delta_i M_i^{-1/4} ( M_i + \beta \Delta_i)^{-3/2} M_i^{-1/4} \Delta_i M_i^{1/2} \right) \; ,
    \]
    where in a slight abuse of notation, we let $M_i^{-1}$ and $(M_i + \beta \Delta)^{-1}$ denote the pseudoinverses of the associated matrices.
    For any $i$, we have that
    \begin{align}
        \tr \left(M_i^{1/2} \Delta_i M_i^{-1/4} ( M_i + \beta \Delta_i)^{-3/2} M_i^{-1/4} \Delta_i M_i^{1/2} \right) & \leq \tr (M_i) \cdot \norm{\Delta_i M_i^{-1/4} ( M_i + \beta \Delta_i)^{-3/2} M_i^{-1/4} \Delta_i}_\op \\
        &\leq C \cdot \tr (M_i) \cdot \gamma^2_i \cdot \norm{M_i^{-1/4} ( M_i + \beta \Delta_i)^{-3/2} M_i^{-1/4}}_{\op} \; ,
    \end{align}
    where the last step follows because $\norm{\Delta_i}_\op \leq \gamma_{i}$ by Lemma~\ref{lem:spectral-projected} and~\eqref{eq:conc-assume}.
   % \begin{align*}
   %      \norm{\Delta_i}_\op \leq \gamma_{i}^2 \leq 2^{-(i + 1)} \; ,
   %  \end{align*}
   By construction, we have that all nonzero eigenvalues of $M_i$ are at least $2^{-i}$, so all nonzero eigenvalues of $M_i + \beta \Delta_i$ are at least $2^{-(i + 1)}$.
    Putting it all together, we obtain that 
    \[
    \tr \left(M_i^{1/2} \Delta_i M_i^{-1/4} ( M_i + \beta \Delta_i)^{-3/2} M_i^{-1/4} \Delta_i M_i^{1/2} \right) \lesssim \tr (M_i) \cdot \gamma_i^2 \cdot 2^{2i} \leq r \cdot 2^{i} \gamma_i^2 \leq \gamma \; ,
    \]
    as claimed.
\end{proof}

\section{Proof of Lemma~\ref{lem:off-diag}}

First, observe that we may disregard the subspace spanned by $\Gamma_{t + 1}$:
\begin{lemma}\label{lem:tiny-subspace}
    Let $\Gbar = \Id - \Gamma_{t + 1}$.
    We have that $\tr (\Gbar \rho \Gbar) = \tr (\Gbar \rho_\diag \Gbar) \defeq c$, and moreover, if we let $\tilde{\rho} = \tfrac{1}{c} \Gbar \rho \Gbar$ and $\tilde{\rho}_\diag = \tfrac{1}{c} \Gbar \rho_\diag \Gbar$, then $F(\tilde{\rho}, \rho) \leq 1 - \gamma / 2$ and $F(\tilde{\rho}_\diag, \rho_\diag) \leq 1 - \gamma / 2$.
\end{lemma}
\begin{proof}
    The first claim follows since $\Pi_i$ and $\Gbar$ all commute.
    The fact that $F(\tilde{\rho}, \rho) \leq 1 - \gamma / 2$ immediately follows from (iv) in Lemma~\ref{lem:projected-norm}, and the last claim follows since the same lemma implies that $\tr (\Gamma_{t + 1} \rho_\diag \Gamma_{t + 1}) \leq \gamma / 2$ as well.
\end{proof}

In light of this, for the rest of the proof, we will assume that $\Gbar = \Id$; in particular, this implies that $\rho$ and $\rho_\diag$ both have minimum eigenvalue at least $2^{-j -1}$ by Lemma~\ref{lem:projected-norm}.
The above lemma implies that this incurs at most an additional additive $\gamma$ to the overall fidelity calculation.
We now establish the following bound on the matrices $E_i$:
\begin{lemma}
    Let $n \gtrsim \tfrac{(d + \log T / \delta) r^2}{\gamma}$, and assume that~\eqref{eq:conc-assume} holds.
    Then, for all $i = 1, \ldots, t$, we have that $\norm{E_i}_\op \leq 2 \gamma_j$.
\end{lemma}
\begin{proof}
    For any $i$, we have that 
    \begin{align*}
        \norm{E_i}_{\op} &= \norm{\Gamma_j \left( \rho - \rho_\diag \right) \Gamma_j}_\op \\
        &\leq \norm{\Gamma_j \rho \Gamma_j - \sigma_j}_\op + \norm{\Gamma_j \rho_\diag \Gamma_j - \sigma_j}_\op \\
        &\leq 2 \norm{\Gamma_j \rho \Gamma_j - \sigma_j}_\op = 2 \gamma_j \; ,
    \end{align*}
    where the third inequality follows because $\Gamma_j \rho_\diag \Gamma_j$ is the projection of $\Gamma_j \rho \Gamma_j$ onto a basis which commutes with $\sigma_j$.
\end{proof}

To prove our overall claim we will proceed via a hybrid argument.
For all $j = 0, \ldots, t$, let 
\[
    \rho^{(j)} = \sum_{i = 1}^j \Pi_i \rho \Pi_i + \Gamma_{j + 1} \rho \Gamma_{j + 1} \; .
\]
Note that by design, we have that $\rho^{(0)} = \rho$ and $\rho^{(t)} = \rho_\diag$.
For these matrices, we show:
\begin{lemma}\label{lem:stagewise-fidelity}
    For all $j = 1, \ldots, t$, we have that 
    \[
        F(\rho^{(j - 1)}, \rho^{(j)}) \geq 1 - 2^{-t - 1} \min (rt, d) \; .
    \]
\end{lemma}

To prove Lemma~\ref{lem:stagewise-fidelity}, we rely on the following fact that bounds the contribution from off-diagonal perturbations to infidelity between two states.

\begin{lemma}
\label{lem:sqrt}
    Let $0 < c_2, c_1 < 1$ satisfy $c_2 \leq c_1 / 10$, and let $M \in \co^{d_1 \times d_1}$ and $N \in \co^{d_2 \times d_2}$ be PSD matrices satisfying $c_1 \Id \preceq M \preceq 4 c_1 \Id$ and $c_2 \Id \preceq N \preceq 2 c_1 \Id$.
    Let $E \in \co^{d_1 \times d_2}$ satisfy $\norm{E}_\op \leq \eta$, and define the matrices
    \begin{equation}
        A = \begin{pmatrix}
            M & E \\
            E^\top & N 
        \end{pmatrix} \; , \; \mbox{and} \; 
        A_\diag = \begin{pmatrix}
            M & 0 \\
            0 & N 
        \end{pmatrix} \; .
    \end{equation}
    Suppose further that $A \succeq 0$, and that $\eta \leq \sqrt{c_1 c_2}$.
    Then
    \begin{equation}
        \tr \left( \sqrt{A_\diag^{1/2} A A_\diag^{1/2}} \right) \geq \tr(A) - c_2 (d_1 + d_2) \; .
    \end{equation}
\end{lemma}
\begin{proof}
    For any $\delta$, let $M_\delta = M - \delta \Id$ and similarly let $N_\delta = N - \delta \Id$.
    we will explicitly construct a matrix $B$ which will be a PSD lower bound for the matrix $A_\diag^{1/2} A A_\diag^{1/2}$.
    Our guess will have the form
    \begin{equation}
        B = \begin{pmatrix}
            M_{c_2} & X \\
            X^\top & N_{c_2} 
        \end{pmatrix} \; ,
    \end{equation}
    for some matrix $X$ we define shortly.
    Note that if we can show that $A_\diag^{1/2} A A_\diag^{1/2} \succeq B^2$, we are done, since by operator monotonicity of the matrix square root, we have that 
    \[
    \tr \left( \sqrt{A_\diag^{1/2} A A_\diag^{1/2}} \right) \geq \tr (B) \geq \tr(A) + \tr(M) - c_2 (d_1 + d_2) =\tr(A) - c_2 (d_1 + d_2) \; ,
    \]
    as claimed.
    
    It remains to demonstrate how to construct such an $X$.
    We choose the following matrix:
    \begin{equation}
        X \defeq \sum_{i = 1}^\infty M_{(c_2 - c_1)}^{-i} \left( M^{1/2} E N^{1/2} \right) N_{(c_2 + c_1)}^{i - 1} \; .
    \end{equation}
We first note that this sum is convergent, since $\norm{M^{-i}_{c_2 - c_1}} \leq \tfrac{1}{(1.9 c_1)^i}$ and $\norm{N_{c_2 + c_1}^i}_\op \leq (1.1 c_1)^i$.
Next, we note that 
\begin{align*}
    M_{c_2} X + X N_{c_2} &= M_{(c_2 - c_1)} X + X N_{(c_2 + c_1)} = M^{1/2} E N^{1/2} \; .
\end{align*}
Therefore,  we have that
\begin{equation}
     B^2 = \begin{pmatrix}
            M_{c_2}^2 + X X^\top & M_{c_2} X + X N_{c_2} \\
            \left( M_{c_2} X + X N_{c_2} \right)^\top & X^\top X + N_{c_2}^2 
        \end{pmatrix} = \begin{pmatrix}
            M_{c_2}^2 + X X^\top & M^{1/2} E N^{1/2} \\
            \left(M^{1/2} E N^{1/2} \right)^\top & X^\top X + N_{c_2}^2 
        \end{pmatrix} \; ,
\end{equation}
At the same time, we have that
\begin{equation}
    A_\diag^{1/2} A A_\diag^{1/2} = 
    \begin{pmatrix}
        M^2 & M^{1/2} E N \\
        \left(M^{1/2} E N^{1/2} \right)^\top & N^2
    \end{pmatrix} \; ,
\end{equation}
    so in particular the off-diagonal block of $B^2$ exactly matches that of $A_\diag^{1/2} A A_\diag^{1/2}$.
    Therefore, 
    \begin{align}
        A_\diag^{1/2} A A_\diag^{1/2} - B^2 &= \begin{pmatrix}
            M^2 - M_{c_2}^2 - XX^\top & \\
            & N^2 - N_{c_2}^2 - X^\top X 
        \end{pmatrix} \\
        &= \begin{pmatrix}
            2 c_2 M - c_2^2 I - X X^\top & \\
            & 2 c_2 N - c_2^2 I - X^\top X 
        \end{pmatrix}
    \end{align}
    Therefore for our candidate to be a valid PSD lower bound, it suffices to show that $c_2^2 I  + XX^\top \preceq 2 c_2 M$ and similarly $c_2^2 I  + X^\top X \preceq 2 c_2 N$.
    Note that for both of these inequalities to be satisfied, it suffices to show that $X^\top X \preceq c_2 N$.
    
    Define the matrix
    \begin{equation}
        Q = \sum_{i = 1}^\infty M_{(c_2 - c_1)}^{-i} \left( M^{1/2} E \right) N_{(c_2 + c_1)}^{i - 1} \; .
    \end{equation}
    Since $N$ and $N_\delta$ commute for all $\delta$, we have that 
    \begin{align*}
        X^\top X = N^{1/2} Q^\top Q N^{1/2} \; .
    \end{align*}
    Additionally, we have that
    \begin{align}
        \norm*{Q}_\op &= \norm*{\sum_{i = 1}^\infty M_{(c_2 - c_1)}^{-i} \left( M^{1/2} E \right) N_{(c_2 + c_1)}^{i - 1}}_\op \leq \sum_{i = 1}^\infty \norm*{ M_{(c_2 - c_1)}^{-i} \left( M^{1/2} E \right) N_{(c_2 + c_1)}^{i - 1}}_\op \\
        &\leq \frac{1.9}{1.1 c_1} \cdot \norm{M^{1/2} E}_\op \sum_{i = 1}^\infty \left( \frac{1.1}{1.9} \right)^i \\
        &\lesssim \frac{\eta}{\sqrt{c_1}} \leq \sqrt{c_2} \; ,
    \end{align}
    by assumption.
    Therefore, we have that $X^\top X \preceq c_2 N$, so $B$ is indeed a valid PSD lower bound on $A_\diag^{1/2} A A_\diag^{1/2}$.
\end{proof}

\begin{proof}[Proof of Lemma~\ref{lem:stagewise-fidelity}]
Fix some $j \in \{ 1, \ldots, t \}$.
Let $B = C_{j + 1}^\top \rho C_{j + 1}$, and define the matrices
\begin{align*}
     A \defeq 
    \begin{pmatrix}
        \rho_{j} & E_j \\
        E_j^\top & B 
    \end{pmatrix} , \; \mbox{and} \; A_\diag = 
    \begin{pmatrix}
        \rho_{j} & \\
         & B 
    \end{pmatrix} \; .
\end{align*}
Then
\[
F(\rho^{(j - 1)}, \rho^{(j)}) = F \left( \begin{pmatrix}
    \tau & \\
    & A
\end{pmatrix},  \begin{pmatrix}
    \tau & \\
    & A_\diag
\end{pmatrix} \right) \; ,
\]
for some shared, unnormalized state $\tau$.
Recall that by Lemma~\ref{lem:eps-bound}, we have that $\rho_j \succeq 2^{-j - 1} \Id$, and that $2^{-t - 1} \Id \preceq B \preceq 2^{-j + 1} \Id$.
Additionally, we have that $\norm{E_j}_\op \leq 2 \gamma_j$.
Therefore the matrices $A$ and $A_\diag$ satisfy the conditions of Lemma~\ref{lem:sqrt} with parameters $c_1 = 2^{-j - 1}$, $c_2 = 2^{-t- 1}$, and $\eta = 2 \gamma_j$.
Note that Lemma~\ref{lem:eps-bound} immediately implies that $\eta \leq \sqrt{c_1 c_2}$, as necessary.
Therefore we have that
\begin{align}
    F \left( \begin{pmatrix}
    \tau & \\
    & A
\end{pmatrix},  \begin{pmatrix}
    \tau & \\
    & A_\diag
\end{pmatrix} \right) &= \tr (\tau) + \tr \left(\sqrt{A_\diag^{1/2} A A_\diag^{1/2}} \right) \\
&\geq \tr(\tau) + \tr(A) - 2^{-t - 1} \left( \sum_{i = 1}^t \mathrm{rank} (\Pi_i) \right) \\
&\geq 1 - 2^{-t - 1} \min (rt, d) \; ,
\end{align}
as claimed.
\end{proof}

We can now finish the proof of Lemma~\ref{lem:off-diag}.
\begin{proof}[Proof of Lemma~\ref{lem:off-diag}]
The desired statement follows by combining Lemma~\ref{lem:tiny-subspace},  Lemma~\ref{lem:stagewise-fidelity} and Corollary~\ref{cor:fidelity-triangle} and redefining $\gamma$ appropriately (scaling down by a logarithmic factor).
\end{proof}

\iffalse

\bibliography{bib}
\bibliographystyle{alpha}

\end{document}
\fi

\section{Conclusion}

In this work we obtained the optimal lower bound of $\Omega(d^3/\eps^2)$ for state tomography in trace distance using adaptive incoherent measurements. Our proof technique is based on setting up a suitable Bayesian problem and arguing that the posterior distribution over the underlying state, conditioned on the measurement outcomes, places negligible mass on the ground truth if $o(d^3/\eps^2)$ measurements are made. This technique is a marked departure from existing approaches to proving lower bounds for adaptive incoherent measurements, which exclusively applied to the \emph{testing} setting. We leave as an open question whether one can refine this lower bound to $\Omega(d r^2/\eps^2)$ in the case where the unknown state is rank-$r$.

Additionally, for the problem of state tomography in infidelity, we gave an \emph{adaptive} algorithm that achieves the optimal rate of $d^3/\gamma$, up to logarithmic factors. By a lower bound of \cite{haah2017sample}, this rate is superior to that of any nonadaptive algorithm. Our algorithm uses logarithmically many rounds of adaptivity, and we conjecture that this much adaptivity is necessary to achieve the optimal rate for incoherent measurements.

% \TODO{mention open problem of getting $dr^2$ lower bound}

\paragraph{Acknowledgments.} The authors would like to thank Steve Flammia and Ryan O'Donnell for coordinating their submission of~\cite{steveryan} with ours. SC and JL would like to thank Jordan Cotler and Hsin-Yuan Huang for illuminating discussions about tomography with incoherent measurements.

\bibliographystyle{alpha}
\bibliography{bib}

\end{document}